\documentclass[twoside,twocolumn,9pt]{article}
\usepackage{extsizes}
\usepackage{apacite}
\usepackage[super,sort&compress,comma]{natbib} 
\usepackage{notoccite}
\usepackage[version=3]{mhchem}
\usepackage[bookmarks=false]{hyperref}
\usepackage[left=1.5cm, right=1.5cm, top=1.785cm, bottom=2.0cm]{geometry}
\usepackage{balance}
\usepackage{subfiles}
\usepackage{caption}
\usepackage{placeins}
\usepackage{comment}
\usepackage{subcaption}
\def\scb{\scriptstyle\bullet}
\usepackage{mathptmx}
\usepackage{sectsty}
\usepackage{graphicx} 
\usepackage{multirow}
\usepackage{lastpage}
\usepackage[format=plain,justification=justified,singlelinecheck=false,font={stretch=1.125,small,sf},labelfont=bf,labelsep=space]{caption}
\usepackage{float}
\usepackage{fancyhdr}
\usepackage{fnpos}
\usepackage[english]{babel}
\addto{\captionsenglish}{%
  
}
\usepackage{array}
\usepackage{droidsans}
\usepackage{charter}
\usepackage[T1]{fontenc}
\usepackage[usenames,dvipsnames]{xcolor}
\usepackage{setspace}
\usepackage[compact]{titlesec}
\usepackage{hyperref}
\usepackage[colorinlistoftodos,prependcaption,textsize=tiny]{todonotes}
\usepackage{epstopdf}

\definecolor{cream}{RGB}{222,217,201}

\begin{document}

\pagestyle{fancy}
\thispagestyle{plain}
\fancypagestyle{plain}{
\renewcommand{\headrulewidth}{0pt}
}

\makeFNbottom
\makeatletter
\renewcommand\LARGE{\@setfontsize\LARGE{15pt}{17}}
\renewcommand\Large{\@setfontsize\Large{12pt}{14}}
\renewcommand\large{\@setfontsize\large{10pt}{12}}
\renewcommand\footnotesize{\@setfontsize\footnotesize{7pt}{10}}
\makeatother

\renewcommand{\thefootnote}{\fnsymbol{footnote}}
\renewcommand\footnoterule{\vspace*{1pt}%
\color{cream}\hrule width 3.5in height 0.4pt \color{black}\vspace*{5pt}} 
\setcounter{secnumdepth}{5}

\makeatletter 
\renewcommand\@biblabel[1]{#1}            
\renewcommand\@makefntext[1]%
{\noindent\makebox[0pt][r]{\@thefnmark\,}#1}
\makeatother 
\renewcommand{\figurename}{\small{Fig.}~}
\sectionfont{\sffamily\Large}
\subsectionfont{\normalsize}
\subsubsectionfont{\bf}
\setstretch{1.125} 
\setlength{\skip\footins}{0.8cm}
\setlength{\footnotesep}{0.25cm}
\setlength{\jot}{10pt}
\titlespacing*{\section}{0pt}{4pt}{4pt}
\titlespacing*{\subsection}{0pt}{15pt}{1pt}

\fancyfoot{}

\fancyhead{}
\renewcommand{\headrulewidth}{0pt} 
\renewcommand{\footrulewidth}{0pt}
\setlength{\arrayrulewidth}{1pt}
\setlength{\columnsep}{6.5mm}
\setlength\bibsep{1pt}

\makeatletter 
\newlength{\figrulesep} 
\setlength{\figrulesep}{0.5\textfloatsep} 

\newcommand{\topfigrule}{\vspace*{-1pt}%
\noindent{\color{cream}\rule[-\figrulesep]{\columnwidth}{1.5pt}} }

\newcommand{\botfigrule}{\vspace*{-2pt}%
\noindent{\color{cream}\rule[\figrulesep]{\columnwidth}{1.5pt}} }

\newcommand{\dblfigrule}{\vspace*{-1pt}%
\noindent{\color{cream}\rule[-\figrulesep]{\textwidth}{1.5pt}} }

\makeatother

\onecolumn

\noindent\LARGE{\textbf{Using small-angle scattering to guide functional magnetic nanoparticle design}} \\

\noindent\large{Dirk Honecker,\textit{$^{a}$} Mathias Bersweiler,\textit{$^{b}$} Sergey Erokhin,\textit{$^{c}$} Dmitry Berkov,\textit{$^{c}$} Karine Chesnel,\textit{$^{d}$} Diego Alba Venero,\textit{$^{a}$} Asma Qdemat,\textit{$^{e}$} Sabrina Disch,\textit{$^{e}$}  Johanna K. Jochum,\textit{$^{f}$} Andreas Michels,\textit{$^{b}$} and Philipp Bender\textit{$^{f}$}} \\

\noindent\normalsize{Magnetic nanoparticles offer unique potential for various technological, biomedical, or environmental applications thanks to the size-, shape- and material-dependent tunability of their magnetic properties. 
To optimize particles for a specific application, it is crucial to interrelate their performance with their structural and magnetic properties.
This review presents the advantages of small-angle X-ray and neutron scattering techniques for achieving a detailed multiscale characterization of magnetic nanoparticles and their ensembles in a mesoscopic size range from 1 to a few hundred nanometers with nanometer resolution.
Both X-rays and neutrons allow the ensemble-averaged determination of structural properties, such as particle morphology or particle arrangement in multilayers and 3D assemblies. 
Additionally, the magnetic scattering contributions enable retrieving the internal magnetization profile of the nanoparticles as well as the inter-particle moment correlations caused by interactions within dense assemblies.
Most measurements are used to determine the time-averaged ensemble properties, in addition advanced small-angle scattering techniques exist that allow accessing particle and spin dynamics on various timescales.
In this review, we focus on conventional small-angle X-ray and neutron scattering (SAXS and SANS), X-ray and neutron reflectometry, gracing-incidence SAXS and SANS, X-ray resonant magnetic scattering, and neutron spin-echo spectroscopy techniques.
For each technique, we provide a general overview, present the latest scientific results, and discuss its strengths as well as sample requirements.
Finally, we give our perspectives on how future small-angle scattering experiments, especially in combination with micromagnetic simulations, could help to optimize the performance of magnetic nanoparticles for specific applications.} 

\twocolumn


\renewcommand*\rmdefault{bch}\normalfont\upshape
\rmfamily
\section*{}
\vspace{-1cm}


\footnotetext{\textit{$^{a}$~ISIS Neutron and Muon Facility, Rutherford Appleton Laboratory, Didcot, OX11 0QX,UK.}}
\footnotetext{\textit{$^{b}$~Department of Physics and Materials Science, University of Luxembourg, 162A~Avenue de la Fa\"iencerie, L-1511 Luxembourg, Grand Duchy of Luxembourg.}}
\footnotetext{\textit{$^{c}$~General Numerics Research Lab, Moritz-von-Rohr-Straße~1A, D-07745 Jena, Germany.}}
\footnotetext{\textit{$^{d}$~Brigham Young University, Department of Physics and Astronomy, Provo, Utah 84602, USA.}}
\footnotetext{\textit{$^{e}$~Universit\"at zu K\"oln, Department f\"ur Chemie, Luxemburger Straße~116, D-50939 K\"oln, Germany.}}
\footnotetext{\textit{$^{f}$~Heinz Maier-Leibnitz Zentrum (MLZ), Technische Universit\"at M\"unchen, Lichtenbergstraße~1, 85748 Garching, Germany.}}








\tableofcontents

\section{Introduction}

\begin{figure*}[ht!]
\centering
\includegraphics[width=1\textwidth]{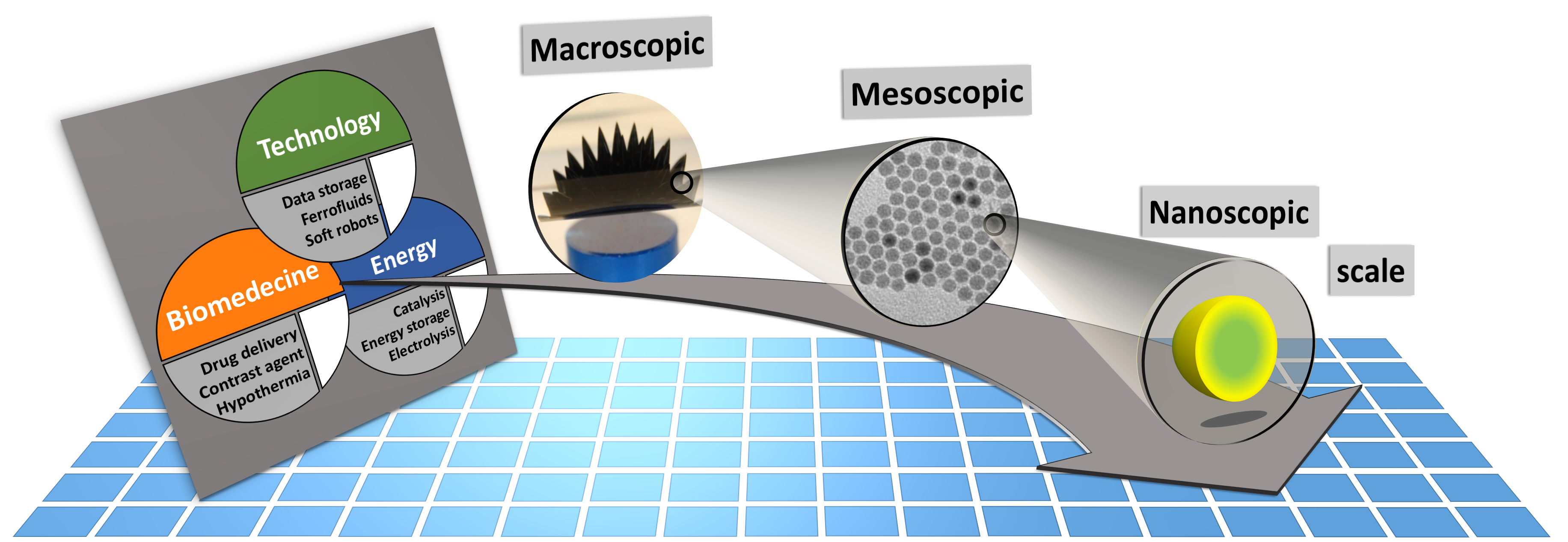}
\caption{The macroscopic properties and consequently the applicability of MNP systems  (e.g. for technological, biomedical, or energy applications) are defined by  their structural arrangement and magnetic interactions on a mesoscopic length scale ranging from a few hundred nanometers down to the individual MNPs. The magnetic properties of the MNPs depend on the particle morphology, chemical composition, and atomic structure.
Small-angle scattering allows to determine the chemical and magnetic structure over the mesoscopic length scale and to connect it with the macroscopic ensemble properties.}
\label{Intro_fig1}
\end{figure*}

Magnetic nanoparticles (MNPs) possess unique tunable properties and can be manipulated (e.g. moved or rotated) by external magnetic fields, making them ideal candidates for various technical, biomedical, and energy applications.
For each application, the employed particles must fulfill specific requirements, especially regarding their magnetic properties.
The magnetism of MNPs can be controlled by various parameters including their morphology, chemical composition, and arrangement.
Some prominent technological applications are their usage in data-storage devices\cite{sun2006recent} or ferrofluids,\cite{odenbach2003ferrofluids} which are stable and dense colloidal suspensions of MNPs used for seals and dampers. 
Ferrofluids rely on superparamagnetic particles stabilized by surfactants that create (reversible) anisotropic aggregates in a magnetic field, whereas the particles envisioned for data-storage devices need strong magnetic anisotropy leading to high coercivities to guarantee thermal stability.
Recently, \citet{Liu2019} demonstrated the reversible paramagnetic-to-ferromagnetic transition by jamming MNPs at oil-water interfaces producing configurable and permanently magnetized emulsion droplets.
This concept can be applied, e.g., to synthesize magneto-responsive nematic liquid crystals.\cite{hahsler2021magnetic}

Typical life science applications, on the other hand, include biosensors,\cite{van2014integrated} magnetic resonance\cite{rumenapp2012magnetic} and particle imaging,\cite{gleich2005tomographic} remote cell control,\cite{dobson2008remote} magnetic drug targeting and delivery,\cite{tietze2015magnetic} as well as magnetic separation\cite{witte2017particle}. 
For each purpose the particle properties need to be optimized: whereas some applications may need superparamagnetic particles (e.g., magnetic resonance imaging), others prefer large particle moments (e.g., biosensors and magnetic separation).
Another prominent biomedical application, which is highly dependent on the used particles is magnetic hyperthermia.\cite{perigo2015fundamentals} 
The working principle of magnetic hyperthermia is to inject MNPs into tumors and heat them by applying alternating magnetic fields to kill the surrounding tumor cells. Commercial heating coil setup provide a fixed frequency (of around 100 - 1000\,kHz) and amplitude (of around 5-20\,mT).
For clinical application, regulatory requirements for biocompatibility restrict the material choice mainly to iron oxides, and thus, only particle morphology and structural arrangement are tuneable parameters to achieve appreciable heating under physiological conditions. 
The unique ability of MNPs to transduce heat in alternating fields on the nanoscale\cite{cazares2019recent} is also utilized for catalysis and energy applications, such as $\mathrm{CO_2}$ hydrogenation\cite{bordet2016magnetically} and electrolysis (i.e. water splitting)\cite{niether2018improved}, in which case the material is not necessarily restricted to iron oxides.

Different advanced synthesis routes exist to design the ideal particles for each application, as we will illustrate exemplarily in the following for magnetic hyperthermia and particle separation.
Generating sufficient heat with alternating fields requires iron oxide MNPs with a significant dynamic susceptibility at high frequencies/low amplitudes.
Recent works indicate that for magnetic hyperthermia large, defect-rich MNPs are great candidates,\cite{lak2020embracing} including so-called nanoflowers\cite{lartigue2012cooperative} and nanocubes with interfacial defects\cite{lak2018fe2+}.
In general, defect-engineering is a promising approach for particle design since it allows manipulating intrinsic MNP properties either by introducing structural defects such as point defects\cite{lappas2019vacancy}, by doping\cite{fernandez2019Cu}, by combining crystalline and amorphous parent material\cite{slimani2021hybrid}, or by controlling twin structures in the nanocrystals\cite{mcdonagh2021new}.
Core-shell particle systems are another particle type explored for magnetic hyperthermia applications\cite{lee2011exchange,lavorato2020origin} as the heating power can be adjusted by tuning the interface coupling\cite{simeonidis2020controlling}.
The control over the magnetic properties via exchange coupling makes core-shell particle systems compelling for various applications in life science\cite{lavorato2021hybrid} and technology\cite{lopez2015applications}.
In addition, particles that are suited for magnetic hyperthermia are commonly also suitable for magnetic particle imaging as the signal is created by alternating magnetic fields.\cite{tay2018magnetic}
In contrast, for magnetic separation, ideal particles have a vanishing intrinsic coercivity but a high static magnetic susceptibility together with a high load capacity and selective and reversible binding affinity.
Particle clusters,\cite{peng2018superparamagnetic}, magnetic microspheres\cite{hafeli2005optical} or large multi-core systems of self-assembled nanoparticles, supraparticles,\cite{guo2013magnetic} are desired as they have large moments and can be thus easily extracted by magnetic gradient fields.
Magnetic separation -- which can be used, e.g. for water purification\cite{iranmanesh2017magnetic} or bioseparation -- often necessitates surface modification to achieve the specific binding of the target compound to the MNPs\cite{ma2007synthesis}.
Particle surfactants can affect the magnetic properties of the functionalized MNP ensembles, which need to be controlled during particle synthesis.\cite{filippousi2014surfactant}
It is worth mentioning that supraparticles are well-suited for magnetic separation and also envisioned for a wide array of applications related to sustainability.\cite{wintzheimer2021}
Nowadays, the synthesis of supraparticles or mesocrystals can be well controlled\cite{park2020strategy} allowing the preparation of particle systems with a wide range of magnetic properties and additional functionalities\cite{haakonsen2021reconfigurable}.

In addition to the synthesis of complex, multifunctional particle systems, the shape and size of the individual nanocrystals can be easily controlled for many materials,\cite{abutalib2021shape} including iron oxides magnetite/maghemite\cite{xie2018shape}, and hematite\cite{meijer2021preparation}.
This allows the preparation of shape anisotropic nanoparticles, which are great candidates for various applications as the magnetic properties can be controlled by particle morphology.\cite{lisjak2018anisotropic}
Elongated ferromagnetic nanoparticles can be applied as nanoprobes for nanorheological approaches when their magnetic moment is preferentially aligned along the long rotation axis due to shape anisotropy.\cite{hess2020scale}
More exotic magnetization states can be found, e.g. in hollow particles\cite{khurshid2021hollow} including  nanorings,\cite{das2020magnetic} nanotubes,\cite{landeros2007reversal} and other shape-anisotropic hollow particles,\cite{niraula2021engineering} or nanodots\cite{goiriena2017magnetization} and nano-octopods\cite{garnero2021single}.
In many cases, the MNPs consist of the typical 3d ferromagnetic elements Fe, Co, Ni, alloys (FePt, FePd) or their oxides, with iron oxides being the most prominent example.
Recently, MNPs of 4f-intermetallic alloys (e.g. $\mathrm{TbCu_2}$ \cite{de2017surfactant} or $\mathrm{GdCu_2}$ \cite{jefremovas2020investigating}) gained interest because their magnetic order can be easily tuned with particle size and microstrain. 
Rare-Earth intermetallics are promising candidates for magnetocaloric applications by tuning the strength of magnetic coupling and modifying the contributions of frustrated and disordered magnetic moments\cite{m2020exploring}.
Furthermore, they excel by a high saturation magnetization and hysteresis-less magnetic response which is desirable for various technological applications.\cite{trepka2020magnetic}
To sum up, a large catalog of different MNP systems exists with unique functionalities tailored with the structural and magnetic properties.
The latter can be achieved by either varying the material, the morphology of the individual nanoparticles, or interparticle interactions.
As illustrated in Fig.~\ref{Intro_fig1}, to optimize a given MNP sample for a specific application, one needs to interrelate their macroscopic properties with their chemical and magnetic nano-/microstructure.

\begin{table*}[ht!]
\caption[]{List of small-angle scattering techniques and their main characteristics: $t_m$ is the typical time needed for one measurement and $t_{exp}$ is the usual time needed for the total experiment.}
\label{Table1}
\resizebox{\textwidth}{!}{%
\begin{tabular}{@{}llllll@{}}
\hline\hline
\textbf{Techniques} & \textbf{Sample and experiment characteristics} &  \textbf{Instruments} & \textbf{Facilities}   & \textbf{Unique features}  \\
\hline\hline

SAXS & $\scb$ $ 20-50\,\mathrm{\mu l}$ (colloids), $ 1-10\,$mg (powders)  & SAXS/WAXS\cite{SAXS_ANSTO} & ANSTO &    \\
+GISAXS & $\scb$ $t_m\approx 1\,$min, $t_{exp}\approx 1-3\,$days &   12-ID\cite{APS_12IDC}
   &  APS   &     \\
+XRR &  $\scb$ scattering power depends on atomic number  & D71\cite{D71}
 & Bessy-II &     \\  
 &  $\scb$ GISAXS/XRR: planar samples $\sim 10\times1\mathrm{mm^2}$  & I22\cite{I22} & Diamond & microfocus option    \\  
 &   & SAXS\cite{SAXSElettra} & Elettra &     \\ 
 &  Retrievable information: & ID02\cite{ID02} & ESRF & USAXS    \\
  & $\scb$ particle size, shape and correlations   & ID12\cite{Goulon2005} & ESRF & XRR, XMCD    \\
 & $\scb$ GISAXS: lateral density fluctuations & P03\cite{P03} & Petra-III &  microfocus   \\
 & $\scb$ XRR: chemical composition profile over thickness  & cSAXS\cite{Liebi2015} & PSI &  coherent diffractive imaging, tensor tomography   \\
& $\scb$ a wide range of laboratory SAXS diffractometers exists & Beamline 1-5\cite{Beamline1-5} & SLAC &       \\
& on the right we give a selection of synchrotron instruments & SIXS & Soleil &       \\  
 &  & SWING\cite{swing} & Soleil &       \\ 
  &  & BL05XU\cite{BL05XU} & SPring-8 &       \\   
 \hline

XRMS & $\scb$ thin samples (e.g. MNPs deposited on Si wafer) & BL29 - BOREAS\cite{Barla2013}  & ALBA        & XMCD/XMLD, extended soft X-ray regime of 80 to 4000 eV     \\
     & $\scb$ beam size $ 0.1\times 0.1\,\mathrm{\mu m^2}$ to $1\times 1\,\mathrm{mm^2}$                                                        & 4.0.2\cite{4-0-2}          & ALS       &  XMCD/XMLD,  400 to 1500 eV     \\
     & $\scb$ $t_m\approx 1-10$\,min, $t_{exp}\approx 1-3\,$ days  &   4-ID-D\cite{4-ID-D}       & APS       &  XMCD/XMLD,  2.7 to 30 keV   \\
     & $\scb$ prior XMCD measurement recommended   & ALICE\cite{Abrudan2015}         & BESSY-II  &   XMCD/XMLD, 8 to 1900 eV     \\
     &  & I10\cite{I10}           & Diamond     &  XMCD/XMLD, 400 to 1600 eV      \\
     &   Retrievable information:  & BM28-XMAS\cite{Xmas}     & ESRF      &  2 to 40 keV       \\
     &   $\scb$ interparticle moment correlations  & ID32\cite{Kummer2016}     & ESRF      & XMCD/XMLD, 400 to 1600 eV photon energy       \\     
     & $\scb$ local disorder and moment fluctuations & I1011\cite{Kowalik2010}         & MAX-II      & XMCD/ XMLD, 200 to 1700 eV     \\
     &  (coherent XRMS)   & 23-ID-1\cite{23-ID-1}       & NSLS-II     & XMCD/XMLD, 250 to 2000 eV   \\
     &   & P09\cite{Strempfer13}           & Petra-III   &  XMCD, variable linear polarisation, 2.7 and 13.7 keV  \\
     &      & EMA\cite{dos_Reis2020}           & SIRIUS      & XMCD, 2.7 – 30 keV  \\
         &     & Sextants\cite{Sextant}      & SOLEIL      & XMCD/XMLD, 50 - 1800 eV  \\
         &    & BL39XU\cite{Maruyama1999}        & SPring-8    &  XMCD/XMLD,  5 - 37 keV \\

\hline  
SANS &  $\scb$ MNP colloids ($V\approx 200-400\,\mathrm{\mu l}$)  & QUOKKA\cite{Wood2018}  & ANSTO  &     \\
+GISANS & $\scb$ MNP powders ($m\approx 100-200\,$mg)   & KWS-1\cite{Feoktystov2015}  & FRM-II  &     \\
 & $\scb$ $t_m\approx 0.1-2$\,h, $t_{exp}\approx 2-5\,$days  & SANS-1\cite{Mulbauer2016} & FRM-II &   TISANE  \\
 & $\scb$ listed instruments allow polarized experiments   & D22\cite{Bender2015,Metwalli2020} & ILL &  TISANE, in-situ SAXS  \\
 & $\scb$ contrast   can be varied by isotope substitution   & D33\cite{Dewhurst2016} & ILL &  monochrome \& time-of-flight mode  \\
 & $\scb$ GISANS: planar samples $\sim 1\times 1\,\mathrm{cm^2}$  & LARMOR\cite{Geerits2019}  & ISIS   & spin-echo techniques + Diffraction     \\
 &  Retrievable information:    & ZOOM\cite{ZOOM_web} & ISIS &  Low-$q$  \\
 &  $\scb$ particle size, shape and correlations   & BL15 TAIKAN\cite{Takata2015} & J-PARC &  High-$q$   \\
 &  $\scb$ magnetization profile within   MNPs  & NG7-SANS\cite{Glinka1998} & NIST &  TISANE  \\
 &   $\scb$ magnetic correlations (structure factor)   & VSANS\cite{VSANS} & NIST &  Low-$q$  \\
 & $\scb$ GISANS: transversal correlations in layer plane &   SANS-1\cite{Kohlbrecher2000} & PSI &     \\ 
 \hline

NR &$\scb$ planar samples $\sim 1\times 1\,\mathrm{cm^2}$   & MARIA\cite{Mattauch_2018} & FRM-II & reflectometry + GISANS  \\
 &  $\scb$ $t_m\approx 1-10\,$min (reflectometry) & SuperADAM\cite{Devishvili2013}  & ILL &  reflectometry + GISANS   \\
 & \ \ $t_m\approx 1-10\,$h (GISANS)  & D17\cite{Saerbeck2018} & ILL & monochrome \& time-of-flight mode + GISANS  \\
 & $\scb$ listed instruments allow polarized experiments  &   CRISP\cite{CRISP_web} and polREF\cite{POLREF} & ISIS &    \\
  &  &  Offspec\cite{Dalgliesh2011} & ISIS  & spin-echo techniques  \\
 & Retrievable information: & BL17 SHARAKU\cite{Masayasu2012} & J-PARC  &     \\
 &   $\scb$ chemical/magnetic composition profile over thickness     & MAGIK\cite{MAGIK} and PBR\cite{PBR} & NIST  &   \\
 &   \ (e.g.   of layered thin films)  & ARMOR\cite{Stahn2016} & PSI  &   \\
 &  &  MAGREF\cite{Lauter2009} & SNS   & reflectometry + GISANS  \\

\hline
NSE & $\scb$ similar sample requirements as for SANS   & RESEDA\cite{Franz2015, Franz2019} & FRM-II &  MIEZE  \\
 & $\scb$ $t_m\approx 2-5\,$h, $t_{exp}\approx 5-10\,$days &   IN15\cite{Farago2015} & ILL &   Low-$q$  \\
 
 & Retrievable information: &   VIN   ROSE\cite{Seto2017} & J-PARC  & Commissioning MIEZE  \\
 & $\scb$ relaxation dynamics in ns to ps-regime &    &   & &  \\
 
\hline

 \hline\hline
\end{tabular}%
}
\end{table*}

To describe the magnetic response of MNP on the macroscopic level, it is often assumed that the individual particles or crystals are single-domains with a homogeneous internal magnetization profile.
Single-domain means that all the atomic magnetic moments $\mathbf{\mu}_a$ within the particle are aligned parallel to each other.\cite{skomski2003nanomagnetics}  
Hence, the total particle moment can be represented by a macrospin $\mathbf{\mu}=\sum\mathbf{\mu}_a$, and thus on the mesoscale MNPs may be regarded macroscopically as simple dipoles. 
The moment $\mathbf{\mu}$ fluctuates with a characteristic relaxation time $\tau$ between energy minima due to thermal activation and it can be distinguished between superparamagnetic (measurement time $t_m >> \tau$) and thermally blocked ($t_m << \tau$) particles.
Only spherical and defect-free particles with diameters below a material-specific single-domain size can be considered model-like, homogeneously magnetized particles.
Real MNP samples always deviate to a certain extent from this oversimplified picture. The intra- and inter-particle magnetization profiles depend on various parameters, including the particle size and shape,\cite{ott2009effects,Gunther2014} as well as dipolar interactions with neighboring particles\cite{morup2010magnetic}. 
Furthermore, structural defects within MNPs are very common,\cite{Nedelkoski2017} especially at the particle surface,\cite{iglesias2004role} which can critically affect their magnetic properties. 
Magnetization or M\"ossbauer spectroscopy measurements indicate the existence of non-negligible  spin disorder in MNPs.\cite{de2017remanence}
The precise determination of the internal 3D magnetization profile remains a key challenge in MNP research and is necessary to fully understand the complex interrelations between their structural and magnetic properties.

Possible approaches to determine the morphology and 3D magnetization profile of nanoscopic systems are electron microscopy-based techniques. 
Transmission or scanning electron microscopy (TEM, SEM) are applied to identify the particle shape and size of MNPs.
With scanning transmission electron microscopy (STEM) a more detailed picture of the atomic structure can be obtained, e.g. resolving anti-phase boundaries within individual MNPs with atomic precision.\cite{Nedelkoski2017}
On the other hand, Lorentz transmission electron microscopy (LTEM) allows detecting the stray field magnetization of MNPs with nanometer-resolution.\cite{phatak2014visualization} 
Higher spatial resolution than LTEM is provided by electron holography, which is sensitive to the entire nanoparticle spin configuration.\cite{gatel2015size}
This technique helped to image the dipolar coupling in planar arrangements of MNPs.\cite{varon2013dipolar} 
Electrons can only probe locally and thin individual structures or arrangement of few particles due to the short penetration lengths. 

To investigate large MNP assemblies or samples with embedded and buried MNPs, we advocate small-angle scattering techniques using either X-rays or neutrons. These techniques cover the technologically relevant mesoscale ($\sim$ 1 - 1000 nm) with nanometer-resolution and enable a structural characterization and the determination of the 3D magnetisation profile of particles and large particle assemblies.
For soft matter systems and structural biology, small-angle scattering is well established as an advanced characterization technique sampling a statistically relevant number of nanoparticles.\cite{Jeffries2021} A general, comprehensive introduction on the non-magnetic theoretical foundations and specifics of both X-ray and neutron small-angle scattering is given in the textbook by \citet{Hamley2021}.
The last decade has seen a continuous drive for new sample environments and continuous development of small-angle scattering based instruments with polarized beam options dedicated for magnetism.
Concerning neutron scattering, the review article by \citet{Muehlbauer2019} provides an overview on dominantly bulk magnetic systems, like magnetic alloys and oxides, noncollinear magnetic structures, skyrmions and flux-line lattices.
This review will close a gap and shows the benefits and recent advances of magnetic small-angle scattering using both neutrons and x-rays to resolve the structural features and essential magnetic information on magnetic nanoparticles and assemblies.
The presented examples demonstrate both the flexibility of the techniques and the breadth of the covered topics e.g. to study the formation of nanoparticles and their assemblies under in-situ conditions, linking the structure with the magnetic response, and probing the magnetization dynamics by covering the relevant timescales. 

In this review, we focus on the following techniques: conventional small-angle X-ray and neutron scattering (SAXS and SANS), X-ray and neutron reflectometry (XRR and NR), grazing-incidence SAXS and SANS (GISAXS and GISANS), X-ray resonant magnetic scattering (XRMS), and neutron spin-echo (NSE) spectroscopy. 
We present for SANS, GISANS, NR, XRMS, and NSE suitable instruments, which can be found at large-scale facilities worldwide.
Table\,1 lists the techniques together with their main characteristics, explaining the kind of samples typically characterized with each given technique, and which information is retrievable.
In the following sections, we present typical examples for each technique regarding MNP characterization and highlight recent outstanding works to show the reader the possibilities to use small-angle scattering to guide functional particle design.
Additionally, new developments regarding the application of micromagnetic simulations for the analysis of magnetic small-angle scattering data are presented.
Finally, we summarize and discuss the topic, and present our perspectives and visions for future research avenues and developments in this field.

\FloatBarrier

\section{Conventional small-angle X-ray and neutron scattering}

\begin{figure}[!ht]
\includegraphics[width=0.5\textwidth]{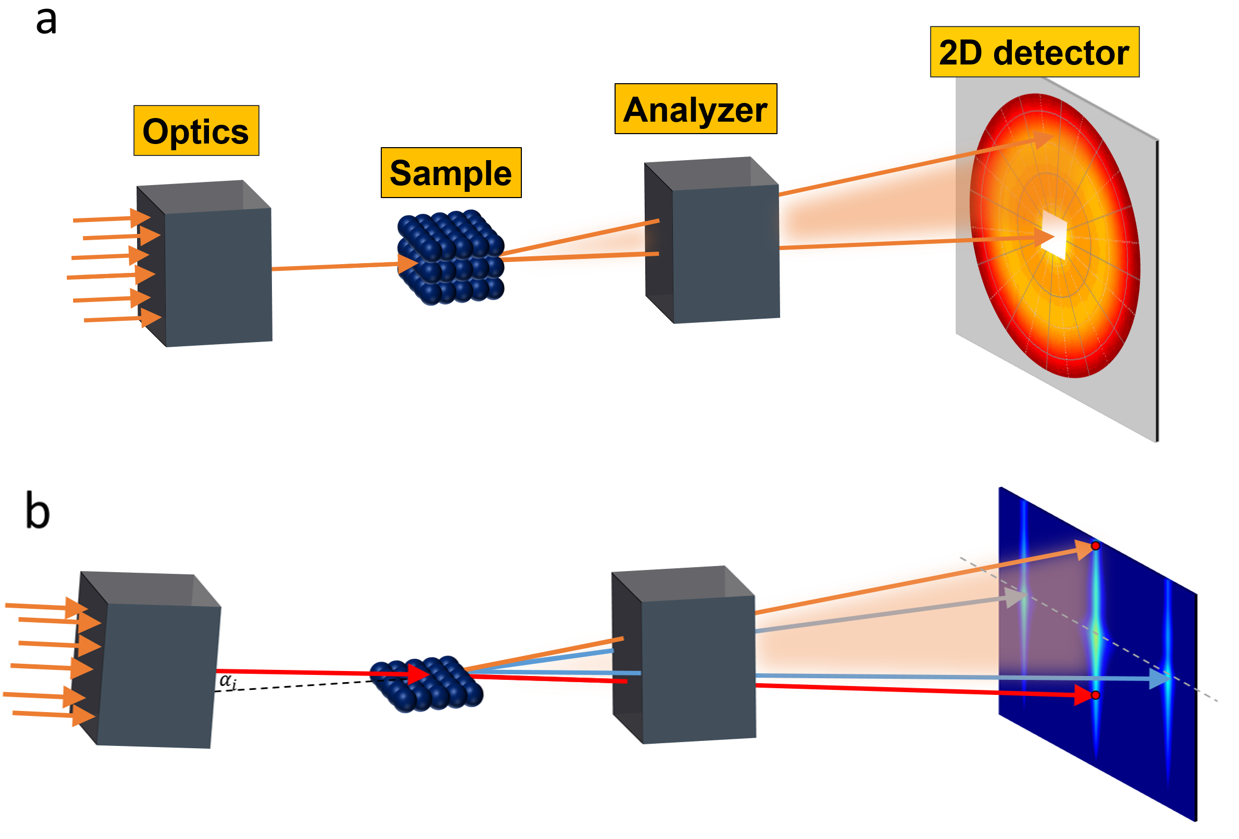}
\captionof{figure}{Small-angle scattering setup in (a) transmission geometry with a position-sensitive detector placed downstream, which is protected by a beamstop (white area) at the center against the direct incoming beam (orange arrows). (b) Reflection geometry with the direct beam (red arrows) grazing the sample under an angle $\alpha_i$. Specular reflection condition (orange arrow) is seen at $\alpha_i=\alpha_f$ above the direct beam on the 2D detector. Interface inhomogeneities give rise to scattering on the vertical incidence line at a different angle than the incident angle. The grazing-incidence small-angle scattering (indicated by the blue arrows) probes the morphology and alignment of nanostructures in the thin film. As an example, the scattering of a square lattice of spheres is shown. 
}
\label{Intro_fig2}
\end{figure} 

\begin{figure}[ht]
    
    \includegraphics[width=0.5\textwidth]{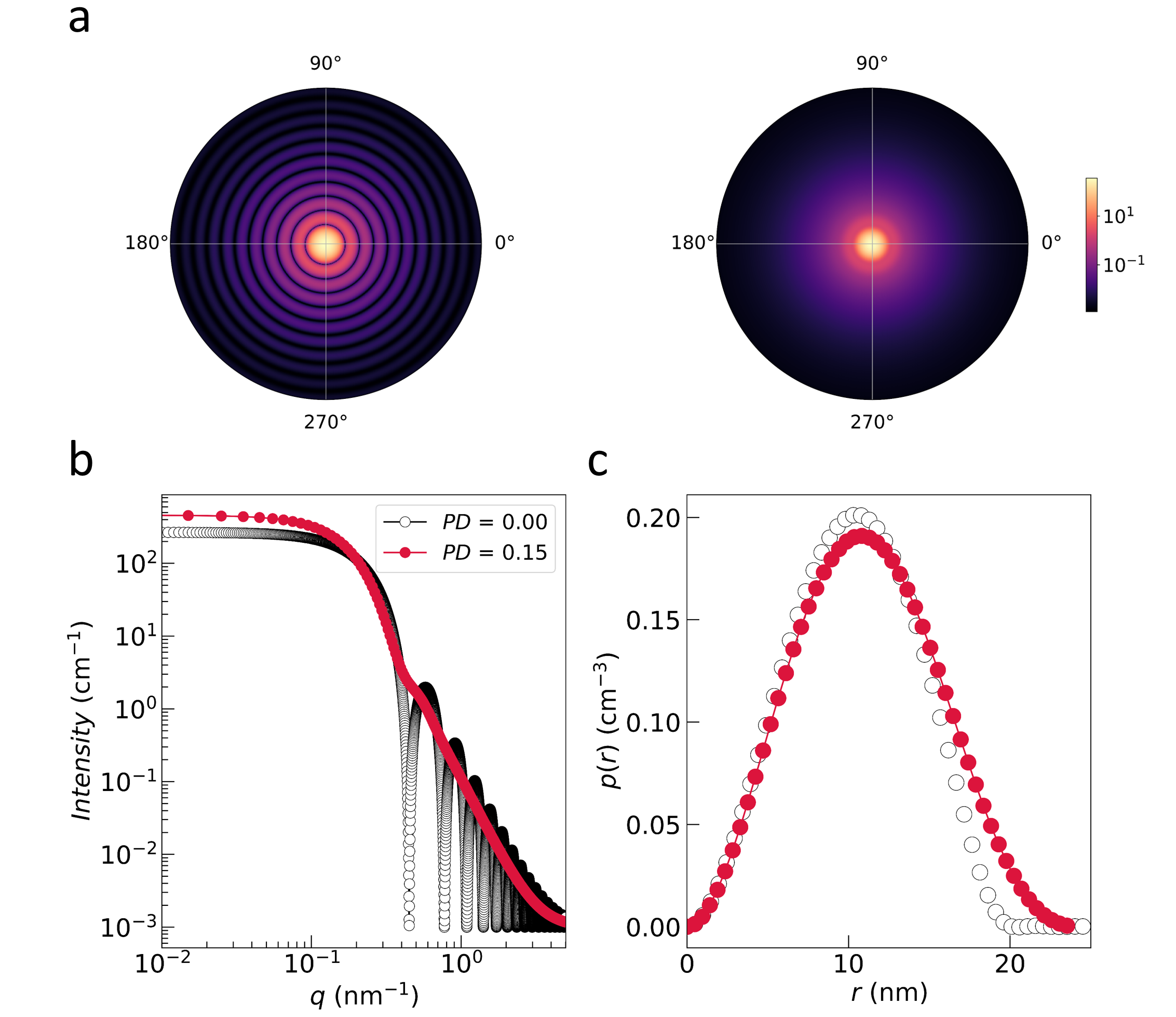}
    \caption{(a) Nuclear 2D scattering pattern of a polydisperse ensemble of spherical, non-interacting particles for the ideal monodisperse case and with an average radius of 10\,nm for a logarithmic size distribution (with width $PD=0.15$). (b) 1D scattering intensity $I(q)$ of the ensemble. The comparison with the form factor $P(q)$ of a single sphere with radius 10\,nm shows how the polydispersity smears out the form factor oscillations at high-$q$, i.e. the Porod range. (c) The corresponding pair distance distribution function $p(r)$ is determined by an inverse Fourier transform of $I(q)$. The data were computed with SasView,\cite{doucet_2021} which provides tools and models to fit data to various particle geometries.}
    \label{fig-SAXS}
\end{figure}

Small-angle scattering allows gathering detailed information on the mesoscopic length scale from about 1 to several hundred nanometers about the chemical composition and magnetization distribution in magnetic nanoparticles.
This allows for example investigating interparticle correlations in aggregates and the formation of superstructures depending on parameters such as particle concentration and external applied magnetic, electric, or flow fields.
The chemical composition and density are associated with the scattering length density (SLD) $\rho(\mathbf{r})$, which for X-rays measures the electron charge density and the isotopic composition for the so-called nuclear scattering of neutrons, respectively.

The measured SAXS and nuclear SANS intensity can be simply written as $I(\mathbf{q})\propto|N(\mathbf{q})|^2$, where $N(\mathbf{q})$ is the Fourier transform of $\rho(\mathbf{r})$ and $\mathbf{q}$ is the scattering vector which is given as the difference between the incoming and scattered wave vector with the magnitude $q=\frac{4\pi}{\lambda} \sin\theta$ for small scattering angles $2\theta$, where $\lambda$ is the wavelength of the incoming radiation. Note that small-angle scattering experiments are conducted either in transmission geometry as illustrated in Fig.~\ref{Intro_fig2}a, or in reflection geometry (Fig.~\ref{Intro_fig2}b).
Often, instead of the 2D scattering pattern, 1D data $I(q)$ are analyzed, e.g. the radial average, which can be written in the case of spherical symmetry as $I(q)\propto\int p(r) \mathrm{sin}(qr)/(qr)\mathrm{d}r$.
Here, $p(r)=r^2C(r)$ is the so-called pair distance distribution function, which is connected to the two-point density-density (Debye) correlation function $C(r)$ of the scattering length density profile.
The correlation function is derived from experimental data by inverse Fourier transforms and provides essential information regarding particle shape, size, density profile, and particle interference, and ---in particular for neutrons---  the magnetization distribution and magnetic inhomogeneities caused by spatial variations in magnetic parameters.\cite{Mettus2015}
Alternatively, in the case of an ensemble of identical, spherical symmetric scatterers, the intensity can be written as the product of the particle form factor $P(q)$ and the structure factor $S(q)$, which can arise from particle interference.\cite{Brunner1997}
For several other particle geometries, analytical functions for $P(q)$ exist. For sufficiently narrow size distributions, the analysis of higher-order form factor oscillation allows resolving finer details of the surface configuration and irregular particle shape, e.g. the degree of truncation and roundness of cuboids.\cite{dresen_2021}
The polydispersity of the nanoparticle ensembles is considered with a corresponding density distribution function.
The relevant structural parameters of the particles and the corresponding distribution function can be determined by model fits of the reciprocal scattering data.
To fit experimental data or to calculate the expected scattering signal of a variety of sample systems, open-source software exists.
Fig.~\ref{fig-SAXS} shows exemplarily the computed data for a polydisperse ensemble of spherical NPs.
For more details regarding the structural characterization of nanoparticle systems by small-angle scattering, we refer to the review article by \citet{li2016small}.

Magnetic small-angle neutron scattering originates from nanoscale variations in magnitude and orientation of the magnetization in a material. 
The dipole-dipole nature implies that magnetic neutron scattering only observes the magnetization component perpendicular to the scattering vector. 
The detected 2D SANS pattern of magnetic samples thus contains additionally to the nuclear scattering contribution a superposition of the Fourier transforms of the three Cartesian coordinates of the 3D magnetization profile $\mathbf{M}(\mathbf{r})=[M_x(\mathbf{r}), M_y(\mathbf{r}), M_z(\mathbf{r})]$.
The weighting of the three contributions of $\mathbf{M}(\mathbf{q})$ depends on the measurement mode (i.e. unpolarized, half-polarized, or fully polarized) as discussed in detail in the review of \citet{muhlbauer2019magnetic}.
Complementary, resonant x-ray magnetic reflectivity relies on measuring the reflected x-ray beam to examine in-depth magnetic profiles of specific magnetic elements in a layered material.\cite{Tonnerre2012} 

Analyzing the spin asymmetry in the reflectivity over an absorption edge for circularly polarized X-ray photons allows obtaining a magnetic scattering contrast that scales with the net magnetic moment.
This particular technique will be introduced in section~\ref{MX}. In this section, we will present an overview of  MNP studies that involved conventional SAXS and SANS.

\subsection{\label{ssec:str}Structural morphology}

SAXS delivers representative statistical data on the morphology of MNPs, particle sizes and size distributions, and the number of individual particles/crystals within aggregates. Studying clusters or aggregates of MNPs is especially relevant with life science applications in mind, as MNPs tend to agglomerate in biological environments such as cells.
This motivated \citet{guibert2015hyperthermia} to study the influence of aggregation of around 12-nm iron oxide MNPs on their magnetic hyperthermia performance via SAXS.
They could correlate an increasing aggregation of the MNPs with a decrease in magnetic heating.
In contrast, the aggregation of smaller iron oxide MNPs enhances magnetic hyperthermia performance.\cite{bender2018influence}
This shows, that depending on the size of the individual MNPs a clustering can either increase or decrease their magnetic heating, which is relevant information regarding a rational particle design for life science applications.
To analyze the structure of MNP aggregates, either model fits or inverse Fourier transforms can be applied.

\begin{figure*}[ht!]
    \centering
    \includegraphics[width=\textwidth]{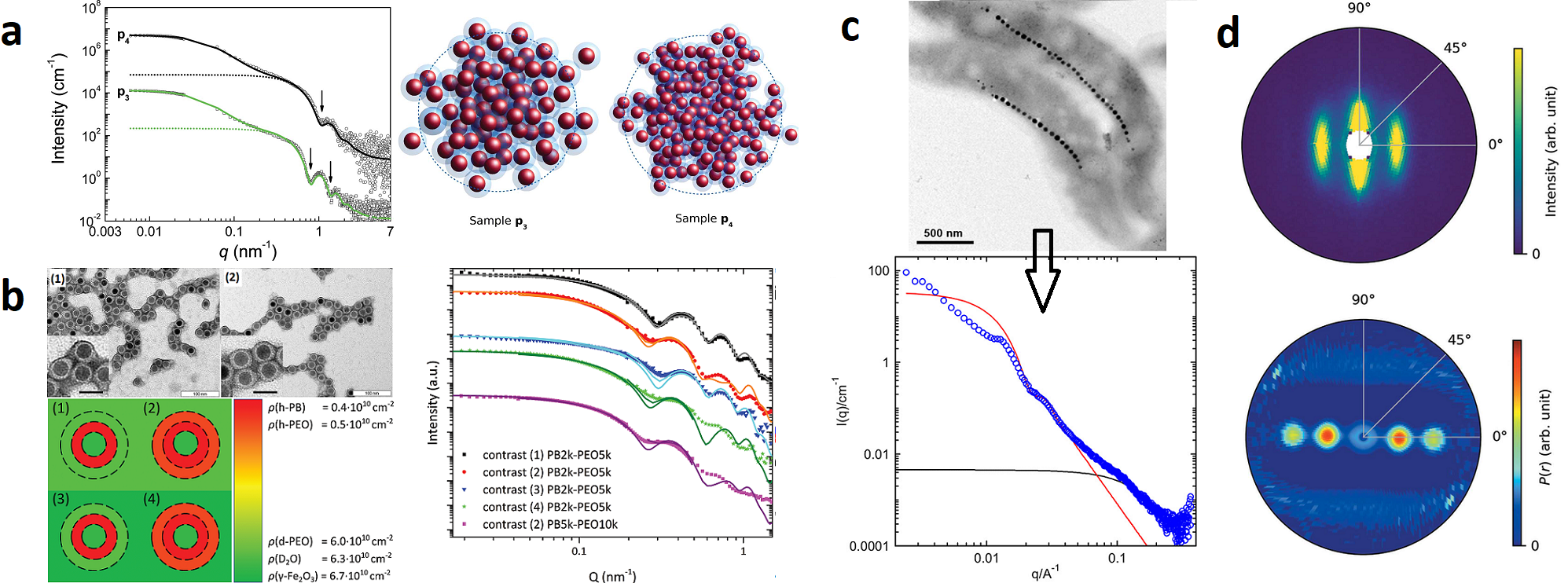}
    \caption{(a) From SAXS measurements of MNP clusters, the individual particle size, cluster size, and compactness can be derived, as exemplarily shown for two different samples in \citet{Szczerba2017}. Reproduced with permission of the International Union of Crystallography. (b)  The structure of iron oxide MNPs encapsulated with a diblock copolymer shell was investigated by SAXS and SANS. By performing a contrast variation, the size of the MNPs and the thickness of the inner and outer shells could be evaluated. Reproduced from \citet{Koll2019} with permission from the Royal Society of Chemistry. (c) In \citet{Rosenfeldt2019} SAXS was employed to monitor the structural changes during the different cultivation stages of magnetotactic bacteria. See the top panel for an electron microscopy image. The 1D SAXS data (bottom panel) were fitted to obtain information about the naturally occurring chains of magnetite particles. (d) The top panel shows the 2D SAXS pattern of a colloidal suspension of magnetotactic bacteria that were aligned by applying an external magnetic field (in the horizontal direction) from which the 2D correlation function (bottom) was extracted by an inverse Fourier transform using a singular value decomposition.\cite{bender2019using} The real-space correlation function reflects the nearest and next-neighbors distance distribution of the linear chain of MNPs and the average size of the isometric MNPs. Reproduced with permission of the International Union of Crystallography.}
    \label{SAS_fig1}
\end{figure*}

Laboratory-based, commercial X-ray facilities provide routine access for small-angle scattering as a primary characterization tool for structural information on nanostructures like the correlations between
the positions of nanoparticles\cite{Bellouard1996} and the packing density\cite{Toro2013}.
\citet{Szczerba2017} for example determined the size and composition of aggregates of iron oxide MNPs, so-called multi-core particles.
The form factor of spheres for the MNPs, a mass fractal structure factor for the aggregates, and log-normal size distributions provided nanoscopic insights into the structure of the multi-core particles (see Fig.~\ref{SAS_fig1}a).
Also, for less defined systems, the aggregate size can be directly inferred via inverse Fourier transform.\cite{bender2018influence}
Alternatively, the form factor of the particles can be a priori assumed, and the size distribution can be extracted by a numerical inversion approach analogous to the inverse Fourier transform. 
This allowed for example the determination of the size distribution of partially aggregated MNPs.\cite{bender2017distribution}

Another type of highly investigated MNPs for biomedical applications are core-shell nanoparticles.
An analysis of inorganic-core/organic-shell systems is often difficult due to the low electron density contrast, e.g. of a hydrated polymer shell against water. 
Exceptionally, SAXS has been utilized to investigate the density profile of highly grafted poly(ethylene glycol)-coated iron oxide MNPs and the temperature-induced contraction of the shell.\cite{Gruenewald2015}
For SANS, contrast variation by hydrogen-deuterium isotope substitution enables to determine the particle morphology and in particular highlight the structure of the surfactant.\cite{Hoell2002,Robbes2012}
For example, \citet{Koll2019} studied the structure of superparamagnetic iron oxide MNPs encapsulated with a highly stable diblock copolymer shell by SANS and SAXS using contrast variation (see Fig.~\ref{SAS_fig1}b).
They could show that their encapsulation process results in a high polymer shell stability, which makes it a great approach for the preparation of MNPs for drug delivery systems.
Contrast variation is also useful to study the influence of surfactant concentration on the stability and aggregation of ferrofluids.\cite{Butter2004,Petrenko2018,Vasilescu2018}
For charge stabilized maghemite MNPs without surfactant, contrast variation allows separating nuclear and magnetic characteristic radii in a water-based medium. 
The surface is covered with around 1/3 of absorbed citrate molecules that mediate the electrostatic repulsion as shown in \citet{Avdeev2009}.
For the above works, the data were in most cases analyzed by model fits.
This is not possible for more complex particles, in this case inverse Fourier transform can be applied to obtain useful information regarding the particle structure of core-shell-type systems.\cite{Bender2017}
For X-rays, contrast variation is achieved by varying the wavelength close to the absorption edge of a particular atom. This anomalous scattering provides element specific sensitivity, e.g. to reveal the chemical composition of 
internal material boundaries Fe$_3$O$_4$ core Mn$_2$O$_3$ shell nanoparticles \cite{Krycka2013}.

Small-angle scattering is a non-destructive method that enables studying a system in its pristine state in comparison to microscopy approaches needing specialized grid preparation, e.g. magnetotactic bacteria in aqueous dispersion.\cite{Rosenfeldt2019} 
These bacteria biomineralize magnetite nanoparticles in specialized organelles (magnetosomes) in a linear arrangement, which enable them to orient along and migrate with the geomagnetic field (see Fig.~\ref{SAS_fig1}c). 
When dispersed in water, the bacteria align along an externally applied magnetic field resulting in anisotropic scattering patterns.
In \citet{bender2019using}, an inverse Fourier transform was applied (here using the singular value decomposition) to extract the underlying 2D correlation function, which nicely reflects the linear chain of aligned magnetosomes (see Fig.~\ref{SAS_fig1}d).
The particle size can be controlled by the cultivation conditions of the magnetotactic bacteria with a typical core radius of 20\,nm and a water-impenetrable organic membrane shell of 3\, nm as seen with neutron contrast variation study.\cite{Hoell2004}  
The MNPs are arranged in a slightly bent chain and the observed misalignment of the particle moment at low magnetic fields is a consequence of magnetic dipolar interactions, magnetic particle anisotropy, and the acting elastic force of the cytoskeleton.\cite{Orue2018} 
The bacteria prove high stability against distortion by magnetic forces as demonstrated with polarized SANS on a fixed freeze-dried powder of the bacteria,\cite{Bender2020a} which make them great candidates for various biomedical\cite{fdez2020magnetotactic}, environmental\cite{wang2020magnetotactic} applications. Recently, for example, their use was suggested for magnonic devices\cite{zingsem2019biologically} and for magnetic actuation\cite{mirkhani2020living}.

\subsection{\label{ssec:mag}Magnetic structure of particles}

Half-polarized SANS (SANSPOL) with an incoming polarized beam, but no analysis of the scattered neutrons, provides an extra contrast given by the interference between nuclear and magnetic scattering and allows filtering backgrounds such as non-magnetic contributions and spin-misalignment scattering arising from moments deviating randomly from the field axis. 
SANSPOL gives access to very weak magnetic contributions, e.g. to infer the existence of a magnetic dead layer in magnetic Fe$_3$O$_4$ glass ceramics,\cite{Wiedenmann2000} to reveal the diffusion mediated Nb enrichment around nanoprecipitates in a metallic matrix,\cite{Wiedenmann2001} and to investigate the magnetization distribution of ferrimagnetic iron oxide compounds with their nuclear SANS contribution often more than an order of magnitude larger than the magnetic signal.
A comprehensive contrast-variation study with varying degree of deuteration of the toluene solvent allows highlighting the penetrability of the surfactant shell surrounding the Co nanoparticles.\cite{Wiedenmann2001}

The magnetic scattering length density determined by polarized SANS is a quantitative measure of the magnetization profile.
For particles below the material-dependent single-domain size, the particle is commonly described as a coherently magnetized ferromagnet (Stoner-Wohlfarth particle).
In this case, small-angle scattering is described in terms of a particle-matrix approach consisting of a form factor describing the shape, size, and orientation of nanoparticles and a structure factor depending on the particle interactions.

\begin{figure*}[ht]
    \includegraphics[width=1\textwidth]{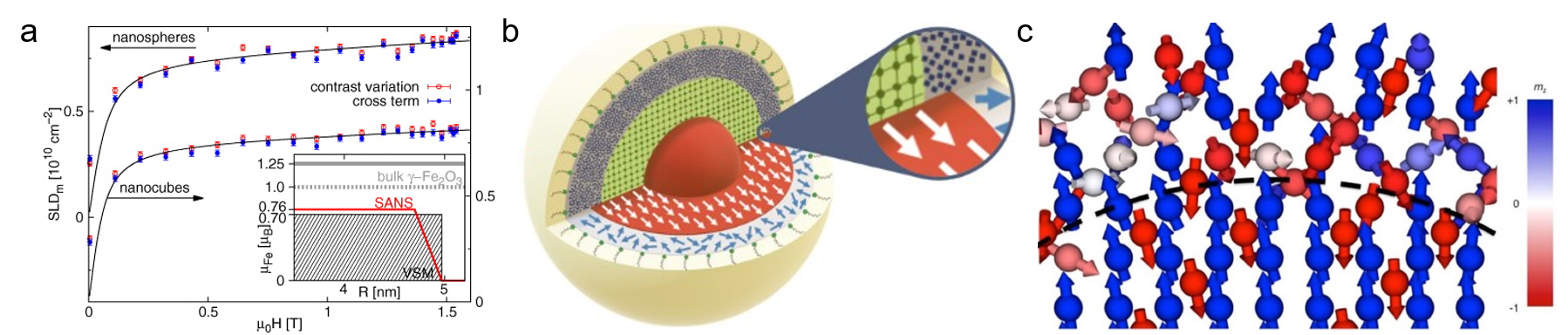}
    \captionof{figure}{(a) Magnetic field variation of the magnetic scattering length density for nanospheres and nanocubes due to particle reorientation described by Langevin behavior (solid lines). The inset displays the magnetization distribution in the nanospheres extracted by SANS compared to the bulk magnetic moment. Reproduced with permission from \citet{Disch2012}. \textcopyright IOP Publishing and Deutsche Physikalische Gesellschaft. All rights reserved.  (b) Representation of the structure (vertical cut) and magnetic (horizontal) morphology of Co ferrite MNP of 14\,nm size. The thickness of the magnetically disordered shell (surface spin disorder indicated by blue arrows) reduces with magnetic field. SANS with polarization analysis (POLARIS) demonstrates that the spins structure on the shell is completely uncorrelated. Adapted from \citet{Zakutna2020} under terms of the CC-BY license. (c) Atomistic simulation of the spin configuration of a Fe$_3$O$_4$ core and Mn-doped ferrite shell with antisymmetric exchange interaction on the octahedral Mn-sites. The dashed line separates the coherently magnetized core from the  shell with local spin frustration leading eventually to a canting of the net magnetization of the nanoparticle. Taken from \citet{Oberdick2018}. Copyright 2018, The Authors, published by Springer Nature. }
    \label{SAS_fig2}
\end{figure*}

In the interior of iron and iron oxide particles, commonly a reduced magnetization compared to bulk material is observed\cite{Butter2004,Avdeev2009} that is mainly associated with the presence of antiphase boundaries\cite{Koehler2021}.
\citet{Disch2012} used SANSPOL to determine the field dependence in iron oxide nanocubes and nanospheres coated with oleic acid (Fig.~\ref{SAS_fig2}a). 
The particle core exhibits a reduced magnetization of 76\% and a gradual demagnetization towards the surface. 
The reduced core magnetization has been associated with microstructural defects within the particle interior resulting in deviations from the perfect ferrimagnetic order,\cite{Wetterskog2013} resulting in spin canting and a random disorder of half the atomic moments as indicated by nuclear resonant scattering\cite{Herlitschke_2016}.
Taking advantage of the spatial sensitivity of SANSPOL, \citet{Zakutna2020} determined that the coherently magnetized core of Co-doped ferrite MNPs grows with the magnetic field (Fig.~\ref{SAS_fig2}b), i.e. inducing magnetic order in the structurally disordered particle surface. 
The magnetic nature of the outer layer was further investigated with SANS with polarization analysis (POLARIS). 
The neutron-spin-resolved measurements indicate uncorrelated, disordered surface spins.
The surface configuration and chemical environment play an important role in the disorder of the surface spins. 
For instance, a coating of iron oxide MNPs with a silica shell enhances the magnetic properties of the surface regions.\cite{Lee2015}


POLARIS also allows observing deviations from single domain behavior to complex spin structures with the presence of ordered misaligned moments from the magnetic field axis. 
The technique measures the neutron spin-state after scattering at the samples allowing to separate magnetic from nuclear scattering using a polarized $^3$He gas cell as analyzing neutron filter. 
For a dense self-assembled face-centered cubic superlattice of 9\,nm Fe$_3$O$_4$ nanoparticles, Krycka \textit{et al.} revealed the existence of considerable, temperature-dependent spin canting of $20-30^{\circ}$ in a 1\,nm surface region.\cite{Krycka2010,Krycka2014} 
A combination of polarized SANS and X-ray magnetic circular dichroism (XMCD) spectroscopy allowed to reveal the evolution of magnetic order in multiphase core-shell nanoparticles. The electronic state and stoichiometry is relatively unaffected with temperature as observed with XMCD, however polarised SANS
proposes magnetic disorder of the oxide shell near the blocking temperature and indicates alignment of the metallic Fe core in the field and reversed magnetization of the surrounding Fe oxide at lower temperatures.\cite{Kons2020} 

Strong spin canting (Fig.~\ref{SAS_fig2}c) is observed in densely packed core-shell Fe$_3$O$_4$@Mn$_x$Fe$_{3-x}$O$_4$.\cite{Oberdick2018}
Atomistic simulations indicate that the effect originates from reduced exchange interaction or Dzyaloshinskii-Moriya interaction between the core and shell phase.

With increasing particle size, inhomogeneous magnetization states occur, not only at the surface but also within the core of MNPs.
In \citet{Bersweiler2019} the purely magnetic SANS signal for Mn-Zn-ferrite samples shows a transition of a nearly homogenous magnetization profile for 28-nm particles to a vortex-like configuration for 38-nm particles.
FePt-core/iron-oxide-shell particles can exhibit a vortex-like intraparticle magnetization configuration reducing dipolar interactions between the particles when no magnetic fields are applied.\cite{Yang2018} 
To identify such complex inhomogeneous conformations, magnetic simulations are a key component. 
The combination of SANS with micromagnetic numerical approaches is discussed in section~\ref{MicroMag}.
 
Multifunctional core-shell nanoparticles enrich the design capabilities for advanced applications.
For instance, \citet{Wang2012} reported the synthesis of colloidal, superparamagnetic particles from iron oxide nanocube. The nanoparticle assemblies exhibit a spherical or cubic shape in a controllable manner by varying the surface tension and the interaction energy between the nanocubes and are highly crystalline as demonstrated with TEM and SAXS (Fig.~\ref{SAS_fig3}a).
In magneto-fluorescent supraparticles, a larger colloidal vesicle encapsulates self-assembled CdSe-CdS core-shell quantum dots with superparamagnetic magnetite MNPs.\cite{Chen2014}
These supraparticles have a size of 100\,nm with an ordered, closed pack core of MNPs surrounded by a shell of randomly distributed quantum dots after thermal annealing.
In \citet{Bender2017} a combination of SAXS and SANS revealed that 9\,nm iron oxide nanoparticle cores accumulate in the surface layer of a 160\,nm polystyrene sphere resulting in a characteristic variation in the scattering length density contrast from core, shell, and solvent. 
A weak dipolar interaction between the particle moments is indicated by the magnetic moment distribution extracted from isothermal magnetization. 
The data analysis used a model-independent indirect transformation method of small-angle scattering and isothermal magnetization data to extract the structural and magnetic distribution.
Such multicore particles are especially interesting for magnetic separation as they have a vanishing coercivity but large effective moments in presence of magnetic fields.

Field‐sensitive ferrogels on the other hand are envisioned for soft actuators.
\citet{Helminger2014} produced biocompatible ferrogels using gelatin gels with embedded magnetite nanoparticles and applied SANS contrast variation experiments to independently explore the nanoparticle packing in the gelatin gel network. 
To optimize the performance of ferrogel-based soft actuators and other functional materials, a good binding between the MNPs and the matrix is necessary as well as a homogeneous MNP distribution.
In \citet{Bonini2008} cobalt-ferrite MNPs were dispersed in polyacrylamide gels and a combination of SAXS and polarized SANS revealed a high quality of their samples.

\subsection{\label{ssec:mag_correlations}Magnetic interparticle correlations}

Small-angle scattering is widely applied to study colloidal stability and ordering in dispersions of nanoparticles. The Brownian motion of magnetic nanoparticles becomes directional with the competition between repulsive stabilizing isotropic forces and the anisotropic dipolar magnetic interactions.
Small-angle scattering with neutron polarisation analysis allowed to measure an alignment in a concentrated cobalt ferrofluid.\cite{Pynn1983}
Even in the absence of particle agglomeration and chaining,  long-range concentration fluctuations along the field indicating a strong anisotropy of the Brownian motion are observed under magnetic field.\cite{Gazeau2002}
Using a combination of scattering methods and reverse Monte Carlo simulations, 
\citet{Nandakumaran2021} explored the chain formation mechanism under magnetic field in solution.
Eventually, dipolar magnetic interactions in ferrofluids induce the formation of superstructure ranging from short-range ordered aggregates via chain-like structures\cite{Barrett2011, Jain2014} to field-induced pseudocrystalline ordering in concentrated ferrofluids\cite{Klokkenburg2007,Fu2016}.

Interparticle forces sensitively affect the rheology of colloidal dispersions.
The magnetoviscous effect, i.e. the strong increase in viscosity with magnetic field, has been investigated with in-situ magnetorheology and small-angle scattering investigating the orientational order.\cite{Krishnamurthy2008,Zakutna2021} 
Shear-thinning behavior is explained by the disruption of the soft aggregates for high enough shear flows.\cite{Pop2006}
Similarly, driven by particle surface charges, a modification of the viscosity can be modified with electric fields in transformer-oil-based ferrofluids.\cite{Rajnak2015}
Magnetic-field oriented aggregates of nanoparticles offer a potential way to obtain strongly anisotropic magnetic properties due to the particle alignment and to cast
reinforced nanocomposite materials with anisotropic mechanical properties.\cite{Robbes2011} 
Magnetic field-induced self-organization is facilitated by large structural and magnetic anisotropies, as shown for elongated hematite nanospindles producing nematically ordered assemblies under a directing static or dynamic field.\cite{Hoffelner_2015}
Anisotropic metallic MNPs as constituents in ferrofluids may result in a
strongly enhanced magnetoviscous effect in comparison to conventional ferrofluids.\cite{Guenther2008,Wu2010}

\begin{figure}[!ht]
    \includegraphics[width=0.49\textwidth]{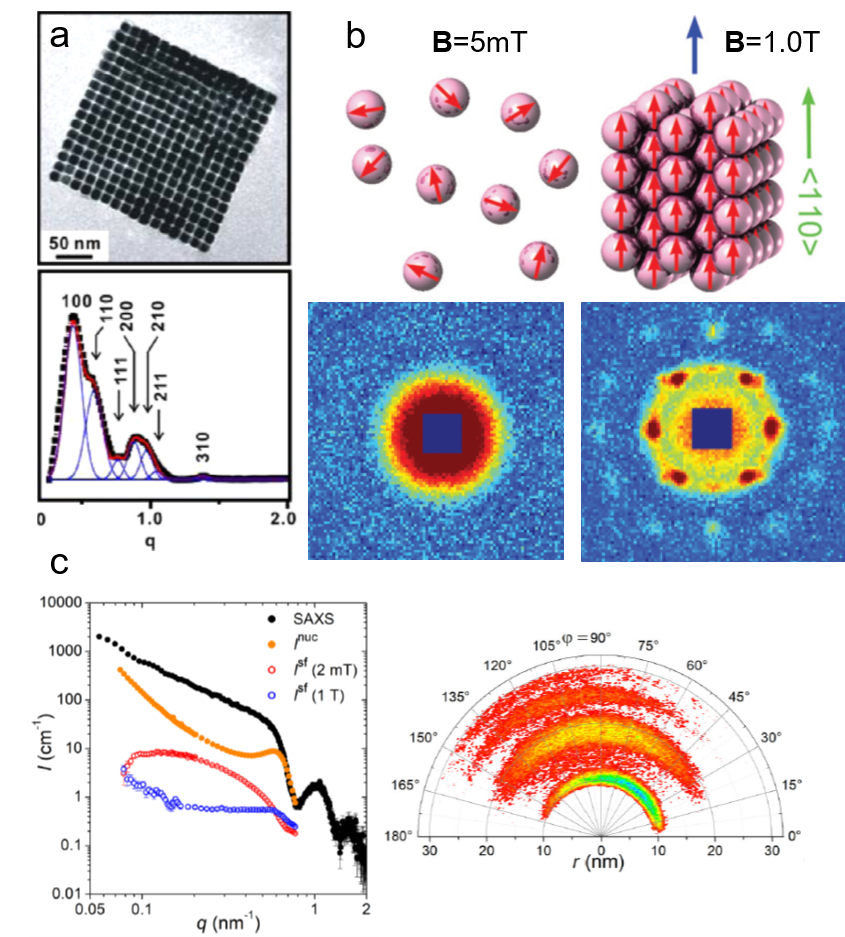}
    \captionof{figure}{(a) The collective properties in superparticles strongly depend on the packing order ranging from amorphous to supercrystalline with well-defined interparticle spacing. Colloidal superparticles of iron oxide nanocubes adopt a simple-cubic superlattice structure. The SAXS pattern consists of the corresponding diffraction peaks. Reprinted with permission from \citet{Wang2012}. Copyright 2012 by the American Chemical Society. (b) SANS patterns of the magnetic-field induced transition from an isotropic, non-ordered colloid to the self-assembly of 3D fcc supercrystals in a 0.1 vol\% dispersion of 17\,nm iron oxide nanoparticles. Taken from \citet{Fu2016} - Published by The Royal Society of Chemistry. (c) SAXS and nuclear SANS scattering reflect the spatial distribution of 10\,nm coated iron oxide nanoparticles in a powder. The feature at $q=0.58\, \mathrm{nm}^{-1}$ is associated with the distance of neighboring particles. The extended high $q$-range for SAXS shows higher-order oscillations of the particle form factor. The scattering at low $q$ reflects interparticle correlations within the clusters. The field-dependent, purely magnetic neutron scattering cross-section $I_{SF}$ resolves the directional correlations between the particle moments. At small magnetic fields, a maximum at $0.12\, \mathrm{nm}^{-1}$ evolves that indicates dipolar interactions in particle clusters up to $70\,\mathrm{nm}$ with a competition between positive and negative moment correlations. (Right) Monte Carlo simulations support the preferential alignment between neighboring moments and dominant anticorrelations for next-nearest moments despite thermal fluctuations. Reprinted figure with permission from \citet{bender2018dipolar}. Copyright 2018 by the
American Physical Society.}
    \label{SAS_fig3}
\end{figure}

Field-dependent SANS is further used to detect collective magnetic correlations among particles in disordered assemblies and ordered particle nanocrystals.\cite{Farrell2006,Sachan2008,Ridier2017}
Small-angle scattering accesses the characteristic length scales connected with interparticle correlations and magnetic interactions between nanoparticles (Fig.~\ref{SAS_fig3}).
In \citet{Dennis2015} the internal magnetic structure of clusters of MNPs could be determined by polarized SANS and connected to their performance for magnetic particle imaging and hyperthermia applications. 
This emphasizes the significant influence of an internal coupling either by dipolar or by exchange interactions on the magnetism of MNP clusters or multi-core particles.
Dextran coated iron oxide multi-core particles, e.g., show a domain structure extending over a stack of parallelepiped structural grains as observed with polarized SANS.\cite{Dennis2009}
Magnetic nanoflowers consisting of sintered iron oxide crystallites are another example of hierarchical nanostructures and are great candidates for magnetic hyperthermia applications thanks to exceptionally high heating rates.
Polarized SANS confirms a preferentially superferromagnetic coupling of the crystallites in a nanoflower resulting essentially in single-domain particles but with a slight spin disorder due to the grain boundaries and other structural defects.\cite{Bender2018}
Furthermore, in the case of a dense powder of such nanoflowers positive correlations between neighboring particle moments were observed creating locally a supraferromagnetic structure.\cite{Bender2020b}
This is in contrast to conventional, spherical MNPs in which case the moments of interacting but superparamagnetic particles tend to align more in an antiferromagnetic-like manner.\cite{bender2018dipolar}
The interparticle magnetic correlations of interacting particles' moments are reflected in the magnetic structure factor, which will deviate from the scattering of the structural arrangement.\cite{Honecker2020}
Interestingly, the interparticle coupling can enhance the magnetic heating of nanoflower samples as shown by \citet{sakellari2016ferrimagnetic} which illustrates the potential of dipolar interactions to (i) drive particle arrangement and to (ii) modify the static and dynamic magnetization behavior of MNP assemblies.
In general, interparticle interactions can be an additional control parameter to produce collective magnetism, and which can be monitored by neutron scattering.
This makes magnetic SANS an invaluable tool to study nanoscopic magnetic correlations in a large variety of MNP samples and other magnetically nanostructured systems.\cite{alba2016magnetic}


\section{Time-resolved in-situ measurements}\label{TR}

\begin{figure*}[ht!]
    \centering
    \includegraphics[width=\textwidth]{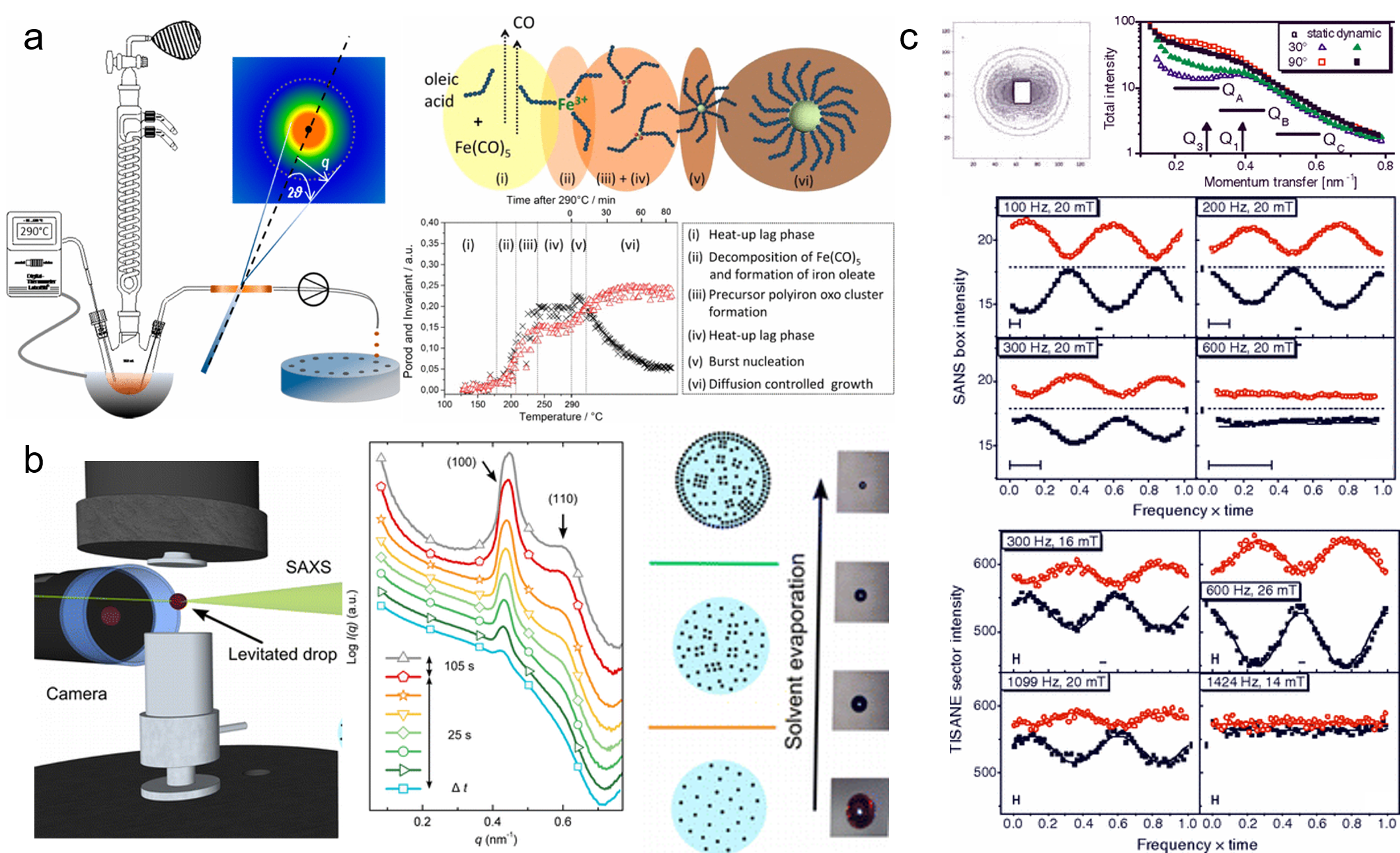}
    \caption{(a) In-situ X-ray scattering allows monitoring the reaction kinetics and precursor states of inorganic and micellar in the synthesis of iron oxide nanoparticles by thermal decomposition. Taken from \citet{Lassenberger2017}. Published under an AuthorChoice License by ACS. (b) Real-time SAXS on a levitated drop allows following the assembly of maghemite nanocubes into mesocrystals. As the solvent evaporates and the drop shrinks, a structure factor appears indicating the particles form clusters and growth of ordered domains both at the air-liquid interface and the interior of the drop. Adapted with permission from \citet{Agthe_2016}. Copyright 2016 American Chemical Society. (c) In concentrated Co ferrofluids, nanoparticle chains are spontaneously formed and order in a local hexagonal arrangement in a static magnetic field, while only partly established in an alternating field. The reversal of magnetic moments is governed by a characteristic Brownian relaxation time of several 100 $\mu s$. The pulsed beam technique TISANE achieves $\mu s$ time resolution allowing to measure the AC frequency dependence of the scattering and overcoming the wavelength smearing for a continuous beam.
    Reprinted figures with permission from \citet{Wiedenmann2006b}. Copyright 2006 by the American Physical Society.}
    \label{TR_fig1}
\end{figure*}

Time-resolved studies with time resolution less than 100~ms are routinely possible on X-ray and neutron beamlines. This ranges from the observation of spontaneous nucleation and growth of particles, further changes due to oxidation over time scales of several days, to the reorientation and switching behavior of the particle moment and the dynamic assembly of superstructures with a magnetic field.
 
A common route to synthesize iron oxide nanoparticles involves precipitating an iron precursor in an alkaline, aqueous solution.\cite{Bee1995,MartinezMera2007} In-situ SAXS helps to identify different reaction pathways that may change with synthesis temperature.\cite{Girod2015} The formation pathway involves intermediate metastable precursors before nucleation and growth of nanoparticles.\cite{Ahn2012, Baumgartner2013,Jensen2014} Ex-situ analysis typically requires sample preparation steps like centrifugation and drying, which potentially can lead to artifacts, e.g. a change in the particle size.\cite{Tuoriniemi2014}
Continuous flow reactors, in which the reagents are pumped and mixed under well-controlled reaction conditions, realize large-scale and reproducible co-precipitation syntheses.\cite{Kumar2012,Simmons2013}
Further, the local position along the reaction tube coincides with the reaction time. This allows observing in-situ the transient reaction states after mixing and to study the growth mechanism\cite{laGrow2019,Besenhard2020} and changes in the magnetic behavior\cite{Milosevic2014}.

In-situ synchrotron measurements are suitable to follow closely the reaction kinetics and precursor state during the synthesis of iron oxide nanoparticles.\cite{Staniuk2014} For instance, \citet{kabelitz2015} followed the formation of maghemite nanoparticles from ferric and ferrous chloride with triethanolamine as a stabilizing agent in an aqueous solution. At various times steps during the synthesis, samples of a few $\mu$l were extracted from the reaction solution and placed in an acoustic levitator to perform X-ray absorption near-edge spectroscopy (XANES) and SAXS. XANES allows determining the oxidation state during the reaction, while the SAXS data detect the growth of particles. The magnetic iron oxide forms rapidly within seconds after mixing the chloride precursor with the NaOH base solution under the abrupt pH change. The co-precipitation is sensitive to local fluctuations of the reaction conditions and affects reproducibility.
In-situ time-resolved, simultaneous SAXS/WAXS studies under supercritical fluid conditions shed light on the synthesis process at 300 bar and above 300$^{\circ}$C allowing to choose suitable residence time to obtain narrow size distributions.\cite{Bremholm2009} Thermal decomposition with high boiling point organic solvents allows synthesizing very monodisperse iron oxide nanoparticles economically in large scale quantities.\cite{sun2002,Park2004}  By continuously sampling the reaction mixture through a X-ray transparent sample chamber (Fig.~\ref{TR_fig1}a), combined SAXS/WAXS experiments resolve the formation of iron oleate complexes, their thermal decomposition to intermediate clusters, and nanoparticle nucleation and growth.\cite{Lassenberger2017} For a reproducible process, an in-depth understanding of the reaction mechanism during each step of nanoparticle formation is needed. The development of in-situ scattering set-ups gives fundamental insights into the nucleation and particle growth kinetics, e.g.
identifying transient amorphous phases and particle aggregation processes in the iron oleate heat-up synthesis, not accounted for in classical description.\cite{Leffler2021}

Post-processing steps may be required for purification and potentially phase transfer to polar solvents via ligand exchange.\cite{Sun2004} Stable, aqueous dispersion of nanoparticles based on amphiphilic polymers are further functionalizable with selected macromolecules,\cite{Yu2006} e.g. for targeted drug delivery. The choice of surfactant can alter structural and magnetic properties.\cite{palma2015} The stability of the aqueous particle dispersion
and absence of interparticle correlations expected after successful phase transfer is easily confirmed using small-angle X-ray scattering.\cite{Zakutna2019} Controlled evaporation of the particle dispersion results in the formation of nanoparticle superlattices\cite{Zeng2004} as discussed further in section ~\ref{GIR}.
To increase the colloidal and chemical stability, magnetic particles can be coated with a protective silica layer, which physically separates the magnetic cores and helps to avoid agglomeration. Time-resolved SAXS has the potential to investigate in-situ the growth kinetics of silica coating on magnetite nanoparticles under various reaction conditions, e.g. the dependence of precursor concentration on the coating process and in particular controlling the shell thickness in relation to the magnetic volume fraction and superparamagnetic relaxation.\cite{Gutsche2014}

Apart from monitoring the growth kinetics of MNPs, time-resolved experiments allow determining the dynamic ordering and relaxation processes of magnetic nanoparticles in magnetic fields, e.g. the formation of chain-like aggregates in a dilute dispersion, which align with an external magnetic field.
The analysis of the scattering cross-section in \citet{Huang2019} indicates that 20\% of the particles form two-bead chains under an external magnetic field. The arrangement is completely reversible when the magnetic field is absent. For Co nanoparticles concentrated up to 6 vol$\%$ dispersed in oil, polarized SANS shows the emergence of sixfold symmetric scattering peaks with a magnetic field indicating reversible pseudocristalline hexagonal order over domains of 100–150$\,$nm, estimated by the width of the correlation peak.\cite{Wiedenmann2003} The order disappears at zero field and the particles arrange in uncorrelated dipolar chains composed of a few particles.
The correlation disappears on the timescale of seconds when the field is switched off.\cite{Wiedenmann2006a} The decay times increase significantly with a field, indicating the stabilizing influence of dipolar interaction on the particle moments relaxation.\cite{Keiderling2007}
Time-resolved unpolarized and polarized small-angle neutron scattering with an AC field and for temperatures down to the freezing temperature of the solvent demonstrates that the magnetic reorientation process of Co and Fe$_3$O$_4$ ferrofluids is composed of moment relaxation characteristic for Brownian rotation of the magnetic cores with finite viscosity or by N{\'e}el type relaxation in the frozen state and a variable volume fraction of arrested, static moments, which can be aligned along with a preferred orientation.\cite{Wiedenmann2008,Wiedenmann2011}
For anisotropic magnetic particles and aggregates, the scattering pattern distortion of a rotating sample in a static magnetic field or a rotating magnetic field allows estimating the rotational diffusion coefficient in the characteristic range up to 1000/s.\cite{Wandersman2009} 
Continuous beam measurements are restricted to AC frequencies below a few hundred Hz. Neutrons passing the sample at a given time arrive at the detector as time-shifted events due to the spread in velocity resulting in a smeared oscillation amplitude of the signal.
Time resolution is hence limited to $1\,$ms.

Regarding SANS, faster relaxation times are accessible with a phase-lock technique called TISANE, which synchronizes microsecond short neutron pulses from high-speed
choppers with a periodic stimulus like oscillatory shear, electric or magnetic fields extending the probed frequency range to the kHz regime (Fig.~\ref{TR_fig1}c).  Similar to stroboscopic data acquisition, the scattering signal is observed over many periods to obtain sufficient counting statistics. A TISANE chopper system has been installed at a few neutron instruments, like SANS-I at FRM-II, D22 at ILL, and NG-7 SANS at NIST.
In a concentrated Co ferrofluid at ambient temperature, field-induced ordering occurs on timescales of $100 \,\mu s$ determined by Brownian rotation, locally ordered domains of $100\,$nm size driven by a dipolar-field governed ordering process are created at a later stage within a few seconds of applying an external magnetic field.\cite{Wiedenmann2006b} The pulsed beam technique has been further used to study the reorientation dynamics of colloidal dispersions of Ni nanorods in oscillating fields.\cite{Bender2015} A recent experiment at NIST investigated hematite nanospindles dispersed in D$_2$O. These anisometric nanoparticles show a significantly different and more complex reorientation behavior.
Sufficiently large nanospindles have a tendency for uniaxial anisotropy overcoming thermal fluctuation of the magnetic moment within the basal plane.\cite{Zakutna2019b}
With the magnetic easy axis fixed in the basal plane, the hematite spindles orient and rotate with their long axis perpendicular to an applied field.\cite{Reufer2010} From an analysis of the frequency-dependent phase delay of the scattering amplitude to the oscillating magnetic field, one can obtain information on the rotational diffusion coefficient of opaque, dense magnetic particle dispersions.\cite{Glinka2020}
The structure-directing influence of static and dynamic magnetic fields\cite{Ahniyaz2007,Leferink2014, Hoffelner_2015} can induce self-assembly of nanocrystals with translational and orientation order that exhibit strongly anisotropic properties that can be used for optical filters, and nanometer-scale viscoelasticity sensors\cite{Eliseev2018}.
Furthermore, doping ferronematic liquid crystals with elongated MNPs such as the hematite spindles is a promising approach to improve their performance, and thus understanding the relaxation dynamics of these composites is crucial for potential applications, e.g., in flat panel displays.\cite{tomavsovivcova2018magnetic}
\FloatBarrier

\section{Reflectometry and grazing-incidence scattering}\label{GIR}

\begin{figure*}[ht]
\centering
\includegraphics[width=\textwidth]{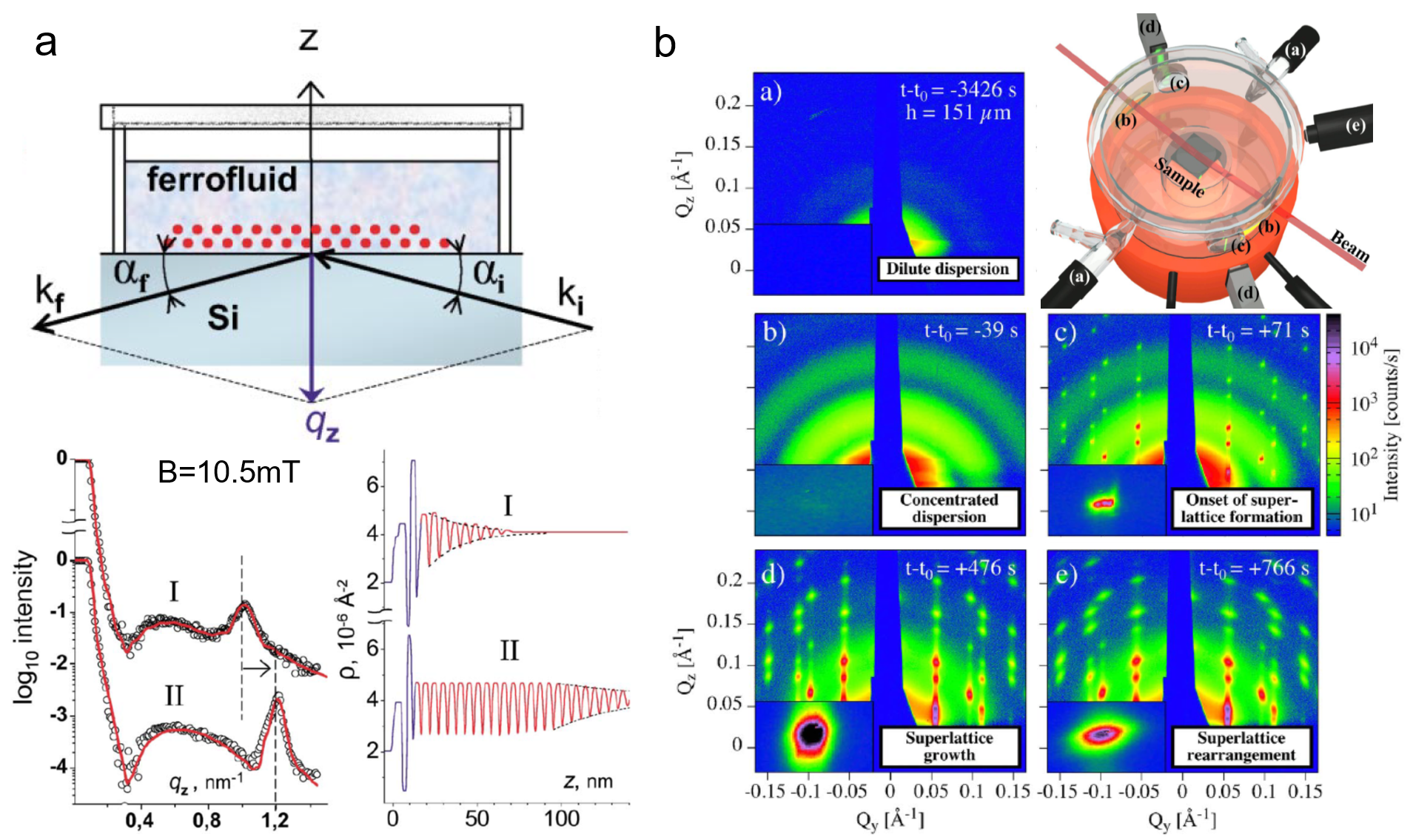}
\caption{(a) Sketch of an experimental specular reflectometry setup with the incident ($k_i$) and reflected ($k_f$) neutron beam and the reflected intensity  versus the momentum transfer $q_z$ normal to the surface after applying a magnetic field of $B=10.5\,$mT perpendicular to the interface for (I) 24 h and (II) 48 hours, respectively.
The scattering length density profile indicates a dense double layer of particles separated by a wetting surfactant layer near the interface (in blue). The Bragg peak (dashed line in the reflectivity) implies the formation and later densification of extended ordered NP layers.  Reprinted figures with permission from \citet{Vorobiev_2004}. Copyright 2004 by the American Physical Society. (b) The evolution of the GISAXS signal highlights the different stages from a dilute dispersion of oleate-capped $\gamma$-Fe$_2$O$_3$ nanosphere in toluene to evaporation-induced self-assembly into superlattices. (Top right) Evaporation cell with gas flow control (a) to adjust evaporation rate, light micrometer (d) for droplet height measurement, and camera (e). Adapted from \citet{Josten_2017}.  Copyright 2018, The Authors, published by Springer Nature. 
}
\label{GIR_fig1}
\end{figure*} 

Nanoparticle assemblies revealing interparticle correlations are typically obtained through either bottom-up self-organization or top-down lithographic techniques. Commonly, a solid substrate supports the sample and provides confinement with a structure-directing influence. The scattering under grazing incidence geometry, or in reflection geometry (Fig.~\ref{Intro_fig2}b), has several advantages for such supported nanostructures. 
First, if the incident angle $\alpha_i$ of the incoming X-ray or neutron beam is small, its large footprint illuminates a relatively large sample area. This results in a much larger scattering volume than in transmission geometry, so that even very thin layers with nanoscale thickness can be studied. Moreover, dynamical scattering effects close to the critical angles for total reflection of the surface and the substrate can be exploited to enhance the scattering intensity and exclusively illuminate the interlayer such that the scattering pattern will be highly sensitive to its structure. 

Reflectometry, off-specular scattering, and grazing-incidence small-angle scattering (GISAS) are closely related experimental techniques exploiting the reflection geometry for the characterization of thin films and interfaces. Excellent literature is available that gives fundamental knowledge about these techniques and the associated scattering theory.\cite{Muller-Buschbaum_2009,Zabel_2008,Wolff_2018} 
Here, we will focus on an overview of the scattering geometries and the different information gained from specular/off-specular reflectometry and grazing-incidence scattering using X-ray and neutron scattering probes.
Scattering in the specular condition (with scattering angle equal to the incidence angle, $\alpha_f = \alpha_i$, Fig.~\ref{GIR_fig1}a) probes the structural and magnetic depth profile of thin films, multilayers, and laterally nanostructured materials. Lateral sample inhomogeneities, such as interface roughness or magnetic domains in the $\mu\,$m regime, give rise to off-specular scattering (with scattering angle unequal to the incidence angle).\cite{Sinha_1988} 
In contrast, lateral structures in the nm length scale result in intensity registered outside the scattering plane, denoted as GISAS (Fig~\ref{Intro_fig2}b). 

Polarized neutron scattering measures the change in the polarization state of the scattered neutron: the sample magnetization parallel to an external magnetic field causes non-spin-flip reflectivity, whereas neutron polarization reversal indicates perpendicular magnetization components, which allows resolving variations of the magnetization vector.\cite{Toperverg_2015}
Polarized neutron reflectometry (PNR) has revealed the dipolar magnetic particle coupling in nanoparticle assemblies observed as domain-like configurations at remanence\cite{Mishra_2015} and through varying layer density\cite{Saini_2020}. 
Polarized GISANS, on the other hand, is an emerging technique that will enhance our understanding of lateral interparticle coupling.\cite{Wolff_2019}

The strong potential of reflectometry and grazing incidence scattering techniques towards depth- and laterally resolved nanostructures finds wide application in the structural and magnetic characterization of MNP assemblies in different dimensions: ranging from nanoparticle monolayers  \cite{Heitsch_2010} through multilayers \cite{Mishra_2012} to highly ordered 3D superstructures such as mesocrystals \cite{Disch_2011,Wetterskog_2016}. 
The following subsections will give an overview of recent achievements using reflectometry (\ref{GIR_depth}) and grazing incidence scattering (\ref{GIR_lateral}), emphasizing also time-resolved studies of in-situ self-organization phenomena.\cite{Theis-Brohl_2020,Josten_2017}

\subsection{Structural and magnetic depth profile}\label{GIR_depth}

Neutron reflectometry has the unique strength to assess depth-resolved structural information, e.g. from buried interfaces, multilayer systems, and nanostructured polymer templates. 
In combination with off-specular scattering, \citet{Lauter_2011} revealed the structure transformation from cylindrical to lamellar structure in nanocomposite films of diblock copolymer and magnetite nanoparticles. 
For assemblies of nanoparticle arrays, the mean distance between the $\mu$m sized supercrystals is accessible using off-specular scattering. \cite{Rucker_2012}
In a combined XRR, GISAXS, and PNR study, \citet{Mishra_2012} analyzed the structural and magnetic ordering of spin-coated nanoparticle films and monolayers. Next to a hexagonal lateral order revealed by GISAXS, a clear modulation of the depth profile was found using XRR, indicating a layered nanoparticle stacking with a linear density gradient between the substrate and top layer. PNR revealed dipolar interparticle coupling and formation of local domains, resembling a soft ferromagnetic state in remanence.  

Neutron reflectometry allows studying the self-organization of nanoparticles from dense ferrofluids at the solid-liquid interface. 
\citet{Vorobiev_2004} revealed the field dependence of the layered structure of ferrofluids near a solid substrate with a dense wetting layer (Fig.~\ref{GIR_fig1}a). The different depth profiles resulting from either static conditions, under shear, and with a magnetic field, are accessible using neutron reflectometry.\cite{Theis-Brohl_2015} Moreover, the ferrofluid properties such as particle surface functionalization strongly affect the first adsorption layers on the solid substrate.\cite{Gapon_2017,Kubovcikova_2017}
\citet{Theis-Brohl_2018}  revealed the influence of different substrates on the wetting layer and subsequent layer formation from ferrofluids and the magnetization depth profile in the obtained assemblies. \citet{Saini_2020} elucidated the impact of a magnetic substrate on the iron oxide nanoparticle layer formation using PNR, highlighting the importance of particle size and the resulting magnetic moment.
\citet{Saini_2019} further used a magnetic field to induce a micro\-shearing effect on small quantities of magnetic polymer nanocomposites that improved the crystallization behavior of nonmagnetic surfactant micelles in water.

\subsection{Lateral and 3D nanostructure}\label{GIR_lateral}

Long-range ordered arrays of nanoparticles, such as mesocrystals or supercrystals, typically assemble as a two-dimensional powder on the substrate with only the substrate normal as the preferred direction. As a result, all (\textit{hkl}) reflections of the sample can be detected using GISAS at the same time.\cite{Disch_2011} 
 In contrast, for an individual, single-crystalline array of nanoparticles, the sample needs to be rotated around the substrate normal to successively fulfill the Bragg condition for different lattice planes.
Lateral structural information can be unambiguously derived from the so-called Yoneda line that emerges at a scattering angle equal to the critical angle of total reflection of the sample ($\alpha_f = \alpha_c$).
For shallow incidence angles $\alpha_i$, both diffraction and refraction processes contribute to the scattering pattern. This leads to two distinct reflections for each lattice plane (\textit{hkl}), which can be indexed by a combination of Bragg's and Snell's laws.\cite{ Disch_2013,Tate_2006} The full GISAXS pattern, including the diffuse scattering resulting from mosaicity, originates from a series of different combinations of reflection and scattering events. For a correct description, multiple scattering effects are considered in the frame of the distorted wave Born approximation (DWBA).\cite{Lazzari_2002,Pospelov_2020} In the case of a rough surface, e.g. for islands of nanoparticle assemblies on the substrate, the condition of an ideally flat sample is no longer valid, and the conventional Born approximation (BA) applies.\cite{Altamura_2012a}

GISAXS is widely applied to the self-organization of nanoparticles into two- and three-dimensional arrangements. For arrangements of semiconductor nanocrystals, GISAXS has been applied to gain insight into oriented attachment,\cite{Choi_2012} the influence of the swelling behavior of the surrounding ligand shell,\cite{Bian_2011} and two-dimensional nanoparticle organization at liquid/air\cite{Geuchies_2020} and fluid/fluid \cite{Balazs_2020} interfaces. 
GISAXS allows comparing quantitatively the quality (substrate coverage, grain size, packing density, and lattice disorder) of FePt nanoparticle monolayers for different Langmuir deposition and spin coating techniques.\cite{Heitsch_2010}
The structure-directing influence of the particle shape has been investigated for cuboidal maghemite nanoparticles, revealing the significant impact of the degree of corner truncation \cite{Disch_2011, Disch_2013} and overall nanoparticle size\cite{Wetterskog_2016} on the formed mesocrystal structures, ranging from body-centered tetragonal to face-centered cubic and simple cubic structure types.
A rich structural diversity of binary nanocrystal superlattices composed of iron oxide and gold nanoparticles and the influence of lattice contraction upon solvent evaporation was reported by \citet{Smith_2009}.
For binary assemblies of CoFe$_2$O$_4$-Fe$_3$O$_4$, SAXS discerned the long-range ordering and high phase purity that results in coherent magnetic switching mediated by the enhanced dipolar coupling.\cite{Yang_2018} \citet{Lu2012} report binary superlattices composed of two different nanoparticle sizes and how excess nanoparticles of one size regime may be expelled and grow separately into locally monodisperse nanoparticle superlattices.

%


Using the high flux of synchrotron beamlines, GISAXS is a suitable tool for the in-situ investigation of the dynamic crystallization processes during the self-organization of nanoparticles. For Au and PbS quantum dots, time-resolved GISAXS enabled a distinction between lateral and three-dimensional growth during the initial stages of superlattice formation as well as overall lattice contraction effects during long-term aging.\cite{Corricelli_2014}
Evaporation of a colloidal nanoparticle dispersion can stimulate spontaneous MNP self-assembly yielding highly ordered and extended nanoparticle superlattices.
The process is determined by a complex interplay between various interactions between particles and substrate, added surfactant molecules, drying kinetics of the solvent, and interfacial energy between surfaces. 
\citet{Siffalovic_2007} revealed the three-phase (liquid-solid-air) drop contact line as the origin of iron oxide nanoparticle self-organization during drop-casting.
The early stages of the self-assembly process of iron oxide nanoparticles in a fast-drying colloidal drop were studied with GISAXS with a temporal resolution down to milliseconds, leading to assemblies with a perfect hexagonal close-packed array within domains with less than 100 nm extension.\cite{Siffalovic2008} 
\citet{Hu_2020} applied vertical scans with transmission SAXS near the liquid interface to identify the region of concentrated NPs and the degree of order above the interface.
\citet{Mishra_2014} reported the influence of wetting vs. dewetting properties of the solvent on the film morphologies of iron oxide nanoparticles on solid substrates. 
\citet{Josten_2017} studied the formation of iron oxide mesocrystals by a drop-casting approach using an evaporation cell designed for in-situ GISAXS with controlled evaporation rates (Fig.~\ref{GIR_fig1}b), identifying four different stages: from a concentrating dispersion to formation, growth, and finally rearrangement of the superlattice. The onset of superlattice formation happens suddenly within seconds indicated by the appearance of sharp structure peaks between two sequential measurements.
A detailed analysis quantifies the ratio of ordered and disordered particle fractions and yields information on the growth kinetics and structural evolution of the superlattice.
Nanoparticle self-organization into 1-, 2-, and 3-dimensional assemblies within the solution is accessible using transmission SAXS.\cite{MehdizadehTaheri_2015} Eliminating the structure-directing influence of the substrate, levitation of drops by ultrasonic waves allows monitoring the two-step assembly of nanoparticles into mesocrystals using time-resolved transmission SAXS (Fig.~\ref{TR_fig1}b). The particles cluster in intermediate, dense, but disordered precursors that rapidly transform into large mesocrystals.\cite{Agthe_2016,Kapuscinski_2020}

Given the well-progressed application of GISAXS to the structural characterization of MNP assemblies, the next important step will be applying grazing-incidence scattering techniques to the understanding of the magnetic morphology in nanoparticle arrangements. 
\citet{Schlage_2012} combined the structural analysis of in-situ grown magnetic antidot arrays using GISAXS with magnetic characterization using nuclear resonance X-ray scattering, revealing the magnetization of the growing iron nanostructure and the impact of an iron oxide capping layer. 
GISANS in combination with polarized neutrons will be a suitable tool to investigate interparticle interactions, such as short-range coupling between nanoparticles in layered assemblies as demonstrated for Co nanoparticles by Theis-Br\"ohl \textit{et al.}\cite{Theis-Brohl_2008,Theis-Brohl_2011}.
\citet{Wolff_2019} applied a magnetic field to induce a micro\-shearing effect on small quantities of magnetic polymer nanocomposites that improved the crystallization behavior of nonmagnetic surfactant micelles in water. The authors suggest a time-resolved and polarized GISANS experiment to elucidate the shear-induced magnetic structure formation and lateral interparticle coupling.
Shear alignment of polymer micelles can serve as a template to impose crystalline order and to fabricate ordered soft magnets.

\FloatBarrier


\section{X-ray magnetic scattering and spectroscopy}\label{MX}

\begin{figure*}[htp!]
\centering
\includegraphics[width=1\textwidth]{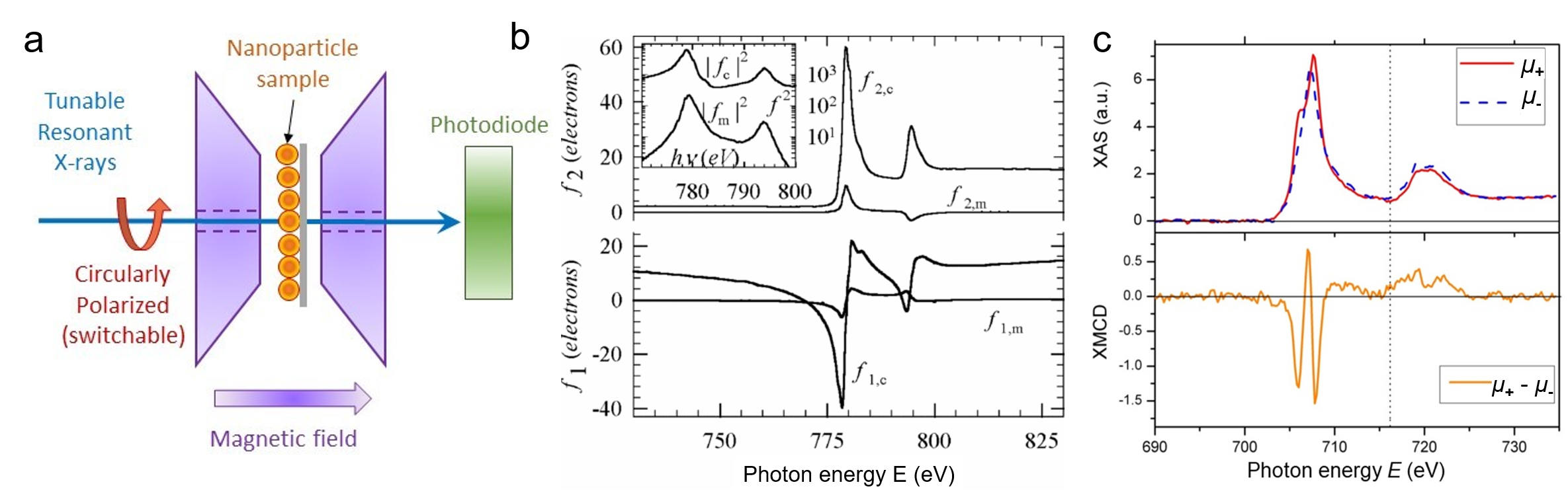}
\caption{XAS and XMCD measurements on MNPs. (a) Sketch of the experimental setup showing the MNP film placed in transmission geometry under an out-of-plane in-situ magnetic field and transmitted light detected downstream on a photodiode. (b) XAS data collected on  9\,nm Co MNPs, displaying the real ($f_1$) and imaginary ($f_2$) parts of the charge ($f_c$) and magnetic ($f_m$) atomic scattering factors. Reprinted figure with permission from \citet{Kortright2005}. Copyright 2005 by the American Physical Society. (c) XAS (top) and associated XMCD (bottom) data collect on 8 nm Fe$_3$O$_4$ MNPs. Adapted from \citet{Cai2014}. Rights managed by AIP Publishing.}

\label{XR_fig1}
\end{figure*}

\begin{figure*}[ht]
\centering
\includegraphics[width=1\textwidth]{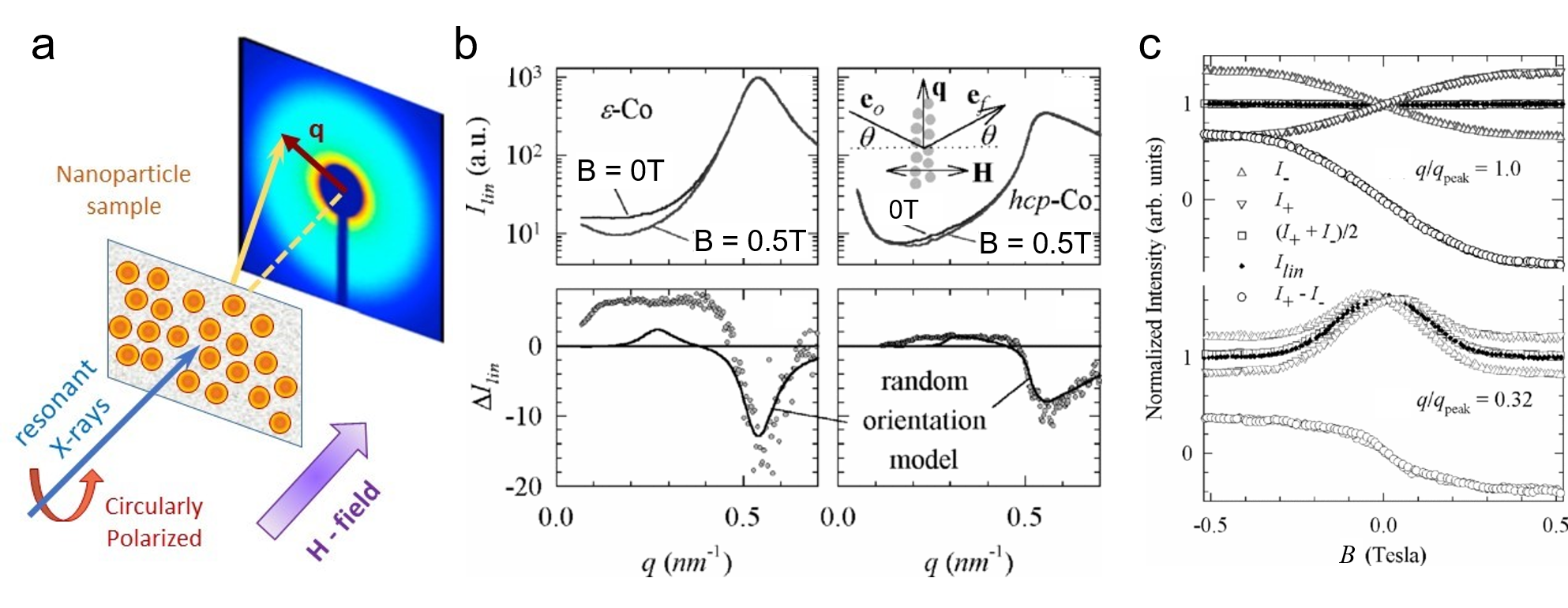}
\caption{XRMS data collected with punctual detection on 9 nm Co MNP monolayers. (a) Sketch of the experimental setup showing the MNP film placed in transmission geometry under an out-of-plane in-situ magnetic field and scattered light detected downstream at transverse scattering vector $q$. (b) (top) sets of $I_{\mathrm{lin}} (q)$ spectra collected at $B=0T$ and $B=0.5\,$T; (bottom) Associated difference $\Delta I_{\mathrm{lin}} (q)$ revealing specific magnetic orders. (c) Magnetic field variation of  $I_{\mathrm{lin}}$, $I_+ (B)$, $I_- (B)$ spectra collected at various $q$ values. Reprinted figure with permission from \citet{Kortright2005}. Copyright 2005 by the American Physical Society.}
\label{XR_fig4}
\end{figure*}

\begin{figure*}[ht]
\centering
\includegraphics[width=\textwidth]{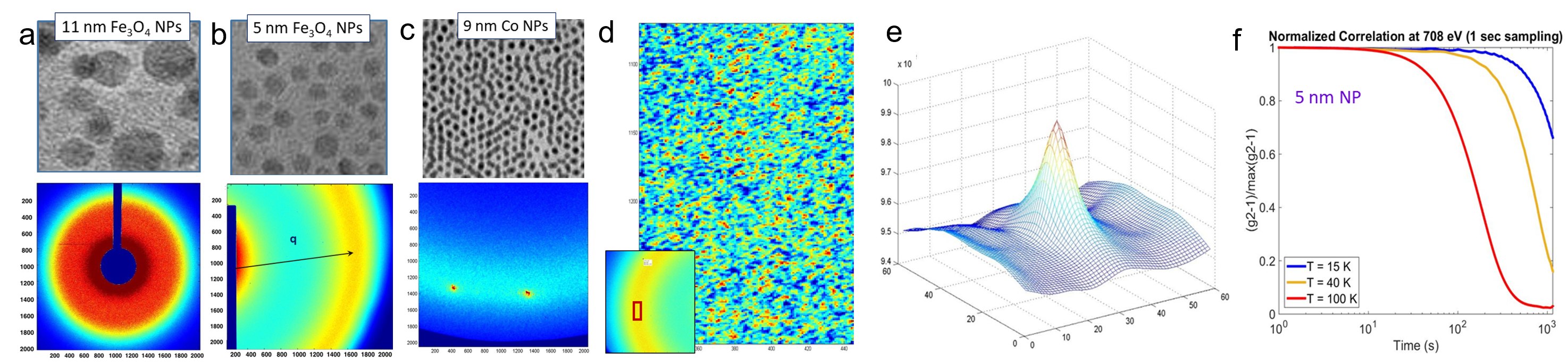}
\caption{Scattering patterns, containing XRMS signal, collected at the Fe-$L_3$ edge on thin layers of MNPs of (a)  11\,nm Fe$_3$O$_4$ MNPs and (b) 5\,nm Fe$_3$O$_4$ MNPs  and (c)  collected at the Co-$L_3$ edge on 9\,nm Co MNPs (Courtesy K. Chesnel). (d) Coherent-XRMS speckle pattern collected on 5\,nm Fe$_3$O$_4$ MNPs at the Fe-$L_3$ edge. This speckle pattern corresponds to the area marked in red of the scattering pattern shown in the inset; (e) Correlation pattern generated by cross-correlating two speckle patterns for estimating correlation coefficient $\rho$. (f)  Normalized time correlation $g^2(t)$ at different temperatures throughout the superparamagnetic transition. (unpublished, courtesy K. Chesnel)}
\label{XR_figcombined2}
\end{figure*}

Synchrotron X-ray radiation offers an advanced tool to probe magnetic correlations in nanostructured materials, such as MNP monolayers and is a unique technique to investigate nanoscale magnetism in the presence of high neutron absorbers such as Sm or Gd.
Due to their high brilliance, synchrotron X-rays enable the detection of scattering signals produced by very thin magnetic layers and low amounts of magnetic materials in a relatively short time, which is often challenging to probe with neutrons. 
In addition, the ability to tune the X-ray energy to specific magnetic resonances provides element selectivity and the magneto-optical contrast necessary to obtain magnetic information. 
Also, switching the X-ray polarization allows separating the magnetic scattering signal from the charge distribution signal, yielding information on the magnetic and structural correlations, separately.
In this section, we will describe two X-ray techniques to study nanostructured magnetic systems, with examples of MNPs. On one hand, X-ray magnetic circular dichroism (XMCD) allows identifying magnetic resonances and extracting information about the spin and orbital moments of a system, but without information on spatial correlations. On the other hand, X-ray resonant magnetic scattering (XRMS) provides spatio-temporal information about nanoscale magnetic correlations.\cite{lovesey1995} 
Also, we will show how \textit{coherent}-XRMS can provide unique information about the local magnetic disorder and the dynamics of fluctuations in MNP assemblies.

\subsection{Magnetic resonances via XMCD}

First predicted and demonstrated in the mid-1980's,\cite{Thole1985,Laan1986} X-ray magnetic circular dichroism (XMCD) exploits the polarization degree of X-rays to probe the magnetic state of matter via absorption spectroscopy.\cite{Galera1995,Goedkoop1997,Stoehr1998}
XMCD is the X-ray equivalent to the magnetic Faraday and Kerr effects observed with visible light in transmission and reflection geometries, respectively.
XMCD is a spectroscopy technique, where the X-ray energy $E$ is scanned across specific absorption edges of the material. 
X-ray absorption (XAS) spectra are recorded at opposite helicities of the circular polarization in order to measure the direction of the atomic magnetic moment relative to the polarization vector of the X-rays. 
The two resulting XAS spectra $\mu_{\pm}(E)$ are then subtracted to estimate the standard dichroic ratio $R_D=\frac{\mu_{+}(E)-\mu_{-}(E)}{\mu_{+}(E)+\mu_{-}(E)} $.

For best magneto-optical contrast, XMCD is typically measured near absorption edges of electronic energy bands for which spin-orbit coupling is present. 
For transition metals, such as Mn, Fe, Co, Ni, XMCD is routinely performed at the $L_2$ and $L_3$  edges where electrons transition from the $2p_{1/2}$ and $2p_{3/2}$ bands, respectively, up to the $3d$ valence band. 
For rare-earth elements, such as Sm, Gd, Tb, Dy, Yb, XMCD is usually carried out at the $M_2$, $M_3$, $M_4$  or $M_5$  edges where electrons transition from the $3p_{1/2}$, $3p_{3/2}$, $3d_{3/2}$ or $3d_{5/2}$ bands, respectively, up to the $4f$ valence band. 
In this band notation, the subscript corresponds to the quantum number $j$, resulting from the combination of the spin $s$ and orbital $l$ angular momenta. 
The strength and shape of the absorption signal $\mu(E)$ (which reflects the electronic density of state in a specific orientation of the magnetic moment associated with $j$) change when the helicity of the polarization is switched.
The resulting dichroic ratio $R_D$ informs on the distribution of magnetic resonance energies, for each magnetic ion and each crystallographic site present in the material. 
Also, information about the spin $s$ and orbital $l$ moments can be extracted by applying the sum rules. \cite{OBrian1994,Piamonteze2009} Combining XAS spectra collected at opposite polarization helicities yields the imaginary part $f_2$ of the charge ($f_c$) and magnetic ($f_m$) atomic scattering factors: $f_{c/m}= f_{1,c/m}+i f_{2,c/m}$ as illustrated in Fig.~\ref{XR_fig1}.
The real part $f_1$ can subsequently be obtained via Kramers-Kronig  transformation.\cite{Kortright2005}

A few XMCD studies of MNPs have been carried out so far, including on Co,\cite{Wiedwald2003,Kortright2005} CoPt,\cite{Imperia2007,Tournus2008} CoFe,\cite{Gao2006} CoFe$_2$O$_4$,\cite{Daffe2018} ZnO,\cite{Chaboy2010}  maghemite $\gamma$-Fe$_2$O$_3$,\cite{Brice2005} magnetite Fe$_3$O$_4$,\cite{Yamasaki2009,Cai2014} hollow MNP,\cite{Bonanni2018} and nanoscale ferrite produced by bacteria and tailoring their magnetic properties by controlled chemical doping\cite{Staniland2008,Coker2009,Byrne2013}. For bimagnetic core-shell systems, element-specific XMCD contributed to determining the interaction between the iron core and spin-canted ferrimagnetic iron oxide shell,\cite{Fauth2004} measuring the magnetothermal behavior,\cite{Jimenez2011} explaining the enhanced anisotropy originating from mixed-oxide interfacial layer\cite{Skoropata2014} and the distribution and coupling of cations\cite{Sartori2019}.
The high sensitivity of XMCD revealed also the induced ferromagnetic order of Ag atoms at the interface of crystalline Fe nanograins and clusters\cite{Alonso_2011} and the polarization of Au nanoparticles by dilute Fe nanoparticles\cite{Venero_2013} in granular thin films.

Fig.~\ref{XR_fig1} shows examples of XAS/XMCD spectra collected at the Co-$L_{2,3}$ edges on 9\,nm Co MNPs\cite{Kortright2005} (Fig.~\ref{XR_fig1}b)  and at the Fe-$L_{3}$ edges on 8\,nm Fe$_3$O$_4$ MNPs\cite{Cai2014} (Fig.~\ref{XR_fig1}c).
The various elements can be separated by identifying the characteristic absorption edges. This allows in particular to investigate the magnetic contributions of mixed ferrites.\cite{Hochepied2001} 
Regarding the data collected at the Fe-L$_3$ edge on Fe$_3$O$_4$ MNPs (Fig.~\ref{XR_fig1}c), the XMCD spectrum exhibits a characteristic W shape with three narrow peaks, which correspond to the various tetrahedral and octahedral sites occupied by the Fe$^{2+}$ and Fe$^{3+}$ ions in the spinel crystallographic structure. 
A quantitative analysis of the XMCD signal using the sum rules indicates that the orbital moment $M_L$ is quenched and that the magnetization is supported by the spin $M_S$ moment. 
The value of $M_S$ is found to be around 2.5 (2.7)\,$\mu_B/$ Fe$_3$O$_4$ at 300\,K (20\,K), which is smaller than the value measured in bulk Fe$_3$O$_4$, suggesting nano-sizing effects and spin canting at the surface of the Fe$_3$O$_4$ MNPs.

These examples demonstrate how XMCD can yield useful quantitative information on the magnetic resonances, magnetic atomic scattering factors $f_{c/m}$, orbital moment  $M_L$, and spin moment  $M_S$, all spatially averaged over the MNP material. 
However, XMCD does not provide any information on spatial correlations. Such information can be obtained locally on discrete nanostructures via imaging, e.g. photoemission electron microscopy with XMCD (XMCD-PEEM) to resolve the magnetization state of single particles,\cite{Fraile2010} or by X-ray scattering techniques over particle ensembles, as discussed in the following.  

It is worth mentioning that XMCD is only present when the material exhibits a non-zero net magnetization (for example, ferromagnetic materials). For anti-ferromagnetic materials, where the net magnetization sums up to zero, the technique of x-ray magnetic linear dichroism (XLMD) is be applied instead of XMCD. 

\subsection{Spatial and temporal correlations via XRMS}

X-ray resonant magnetic scattering (XRMS) exploits the X-ray polarization to probe spatial magnetic correlations in the matter via scattering.\cite{Blume1985,Hannon1988}
To optimize the magneto-optical contrast, the energy of the X-rays needs to be finely tuned to resonance edges of the magnetic element(s), which are often shifted relative to the theoretical tabulated edges, due to the electronic state of the excited atom in the local chemical and magnetic environments. Additionally, magnetic resonances may be sharper than electronic resonances. For these reasons, and given potential energy shifts in beamline calibrations, it is generally recommended to perform XMCD measurements prior to XRMS  identifing the exact magnetic resonance edges to optimize the magneto-optical contrast.  

Unlike for XMCD, where integrated absorption spectra are measured either as drain current from the sample, total fluorescence yield or as transmitted photon intensity (see Fig.~\ref{XR_fig1}a), the XRMS intensity is recorded as a function of the scattering momentum transfer $\mathbf{q}$ (see Fig.~\ref{XR_fig4}a) and necessitates scanning the photo-detector spatially or using two dimensional detection downstream.

Historically, XRMS was first used in the hard X-ray regime, using L-absorption edges of rare-earth elements, for example to characterise magnetic nanostructures in granular amorphous GdFe alloys.\cite{Fischer1998} It is also worth mentioning that with sufficient magnetic contrast, non-resonant magnetic scattering of hard X-rays  may be used to obtain information about existing atomic magnetic ordering\cite{Barbier2004}. One advantage is the deep penetration of hard X-rays compared to soft X-rays, allowing to study thicker, even $\mu m$ thick materials with XRMS in transmission geometry.

However, XRMS is nowadays mostly used in the soft X-ray range to study nanoscale magnetic structures, such as magnetic multilayers,\cite{Tonnerre1995,Hase2000,Chesnel2001,Spezzani2002,Marrows2005,Valvidares2008} magnetic domains in  ferromagnetic thin films, \cite{Dudzik2000,Kortright2001,Kortright2002,Nefedov2004,Miguel2006,Asti2007,Fin2015,Lamirand2017,Flewett2017} noncollinear spin textures in magnetic thin films,\cite{Meyerheim2009,Langner2014,Zhang2018,Legrand2018,Chauleau2018,Diaz2019}  magnetic ordering in antiferromagnetic system\cite{Hellwig2007} and exchange-coupled magnetic layers \cite{Radu2006,Chesnel2008b}. When tuned to the Co and Fe-$L_3$ edges, the X-ray wavelength is around $\lambda = 1.6\,$nm and 1.75\,nm, respectively (soft X-rays). These wavelengths are perfectly suited for the study of MNP assemblies, as it gives access to spatial ranges from a few nanometers up to about 100\,nm depending on the resolution and the angular extent of the detector. So far, only a few studies have utilized XRMS to probe magnetic nano-objects, such as Co/Pt nanowire lattices,\cite{Chesnel2002a} FePd single nanowires,\cite{Chesnel2002b} Co MNPs,\cite{Kortright2005} patterned nanomagnet arrays,\cite{Morgan2012} and Fe$_3$O$_4$ MNPs,\cite{Chesnel2018} as illustrated in  Fig.~\ref{XR_fig4} and Fig.~\ref{XR_figcombined2}a-c. 
For all these systems, the X-ray energy is tuned to either Co-$L_{2,3}$ or Fe-$L_{2,3}$ edges in the soft X-ray range. Because these materials are nano-structured, soft x-rays are well suited to access their inherent nanoscale magnetic correlations.

For isotropic materials that do not show any preferential direction, one can use a point detector, typically a photodiode mounted on a rotating arm that may allow scanning the scattering angle $2\theta$ through a wide range from 0 up to $90^{\circ}$ similar to powder X-ray diffraction (XRD). However, most of the XRMS signal is often concentrated in the small angle region. 
An example of scan in the small-angle regime is illustrated in Fig.~\ref{XR_fig4}b, where the XRMS signal is recorded at the Co-$L_3$ edge on 9\,nm Co MNPs.\cite{Kortright2005} 

To extract magnetic correlations, $I (q)$ data sets are recorded at left and right circular ($I_{\pm}$) as well as linear ($I_{\mathrm{lin}}$) X-ray polarization, and at various magnetic field values $B$. 
Using a point detector allows to study the dependence of the scattered intensity $I(B)$ with the magnetic field $B$ at a fixed $q$ value, as illustrated in Fig.~\ref{XR_fig4}c. 

The two-dimensional detection mode, similar to single crystal detection, provides 2D lateral spatial resolution, particularly informative for non-isotropic nanostructured materials. 
In this case, a two-dimensional detector such as a CCD camera is placed either directly downstream if the material is probed in transmission (see Fig.~\ref{XR_fig1}a), or positioned at some angle if the material is probed in reflection (see Fig.~\ref{Intro_fig2}b). 
When done in transmission, the probed scattering angle $2\theta$ is typically limited to small values, making XRMS fall in the small-angle X-ray scattering (SAXS) category, analog to small-angle neutron scattering (SANS).

Fig.~\ref{XR_figcombined2} shows examples of 2D XRMS patterns collected on self-assemblies of Fe$_3$O$_4$ MNPs with the energy $E$ tuned to the first magnetic peak within the Fe-$L_3$ edge (Fig.~\ref{XR_figcombined2}a,b), and on assemblies of Co MNPs with energy tuned to the Co-$L_3$ edge (Fig.~\ref{XR_figcombined2}c). Because the MNP materials are nanostructured, the scattering signal is actually a combination of conventional (Thomson) charge scattering (induced by structural correlations) and magnetic scattering (induced by magnetic correlations). The shape of the scattering signal in (Fig.~\ref{XR_figcombined2}a,b) is an isotropic ring, due to the fact that the scattering signal is collected over a large portion of the sample, averaging many various orientations of short-range MNP assemblies. The radius of the ring informs on the average inter-particle distance, whereas its width provides information on correlation lengths. On the other hand, the scattering pattern in Fig.~\ref{XR_figcombined2}c shows a set of local peaks, revealing in this case long-range ordered lattices of NPs.  To extract information about magnetic periodicities and magnetic correlation lengths, some further steps are necessary, exploiting the dependence of the signal with varying magnetic field as well as polarization dependence, as explained below. 

The scattering intensity $I (q)$ collected in opposite helicities of circular polarization can be decomposed as a mix of charge and magnetic scattering amplitudes, $A_c$ and $A_m$, and be written as follows \cite{Kortright2001,Chesnel2018} $I_{\pm} (q)=|A_c+A_m|^2=|A_c|^2 \pm (A_c^* A_m + A_m^* A_c) + |A_m|^2 $. In this expression, the respective charge/magnetic scattering amplitudes $A_{c/m}=f_{c/m}\,s_{c/m}(q)$ are made of the atomic scattering factor $f_{c/m}$ and the structure factor $s_{c/m}(q)$ that contains information on the charge/magnetic structure of the material. The factor $s_{c/m}(q)$ is essentially the Fourier Transform of the charge/magnetic density functions.
 
To access the information on magnetic correlations specifically, one method consists in looking at the reconstructed linear intensity $ I_{\mathrm{lin}} (q)=   I_+ + I_- =|A_c|^2 + |A_m|^2$, and following its field dependence by collecting data at different field values, typically at $B = 0$ and at a saturating field $B_{\mathrm{max}}$. If the charge component $A_c$ is not field-dependent, one can access the magnetic component $A_m$ by measuring the difference\cite{Kortright2005}$\Delta I_{\mathrm{lin}} (q)=I_{\mathrm{lin}}(B_{\mathrm{max}})-I_{\mathrm{lin}}(B=0)=\Delta |A_m|^2 =\Delta |s_m (q)|^2$. This method, illustrated in Fig.~\ref{XR_fig4}b,c, only works if the magnetic scattering is not negligible compared to charge scattering so the variation can be effectively measured.

If $|A_m|^2\ll|A_c|^2$, another, a more suited approach consists in exploiting the dichroic difference and calculating the magnetic ratio, defined as $R_M==\frac{I_+ -I_-}{\sqrt{I_+ -I_-}}$. In the small magnetic scattering approximation, one can show that $R_M\approx \frac{2Re(A_c A_m)}{|A_c|^2}\propto  |A_m| \propto  |s_m (q)|$, thus giving access the magnetic scattering amplitude $|A_m|$ directly rather than its square $|A_m|^2$\cite{Chesnel2018}.

The $\mathbf{q}$-dependence of the magnetic scattering allows obtaining information on spatial magnetic correlations, such as inter-particle ferromagnetic and antiferromagnetic couplings, which correspond to a preferential parallel and antiparallel alignment of magnetic particle moments, respectively.\cite{Rackham2019} 


Finally, $coherent$ X-ray resonant magnetic scattering (C-XRMS) makes use of the high brilliance and coherence of synchrotron radiation, which allows producing wavefronts highly correlated in space and in time. 
Coherence may be achieved in two ways: longitudinally (temporal), by selecting a monochromatic wavelength band, or laterally (spatial), via pinhole filtering and reducing the source size. 
Under coherent illumination, the interference between scattered beams from different parts of the  material produces a speckle pattern as illustrated in Fig.~\ref{XR_figcombined2}d. 
The particular position and shape of the speckle spots reflect the local charge distribution and magnetic structure in the material. 
The speckle pattern is a unique fingerprint of the nanoscale correlations, e.g. the size and distribution of the magnetic domain configuration.\cite{Chesnel2016} 
C-XRMS provides information on short-range inter-particle ordering as well as on slow dynamical behavior in MNP assemblies.\cite{Chesnel2008a}

In the cross-correlation process, two speckle patterns $A$ and $B$ are compared by superimposing them with some gradual lateral shift and multiplying them pixel-by-pixel. 
This operation (denoted by the symbol $\times$) produces a correlation pattern, such as the one shown in Fig.~\ref{XR_figcombined2}e. 
By integrating the signal under the peak ($\sum$ operation) and normalizing by the auto-correlations of the two speckle patterns, one obtains a normalized correlation coefficient\cite{Pierce2005,Chesnel2012}$\rho=\frac{\sum (A \times B)}{\sqrt{\sum (A\times A) \sum (B \times B)}}$ which varies between $0$ (no correlation) and $100\%$ (same exact pattern).

The coefficient $\rho$ can be estimated over the entire speckle pattern, or on specific regions and $\mathbf{q}$ values, e.g. characteristic for short-range magnetic correlations in the plane of a monolayer of MNPs.\cite{Chesnel2011} 
The dependence of $\rho$ on an applied magnetic field and temperature shows the evolution of these magnetic correlations throughout a magnetic hysteresis and over-phase transitions.
Similar to comparing the correlation coefficient $\rho$ for two images $A$ and $B$ collected at different times, another, more common approach, known as X-ray photon correlation spectroscopy (XPCS), consists in estimating the correlation coefficient\cite{Gruebel2004,Shpyrko2014} $g^2(t)=\frac{\langle I_A(t)I_B(t+\Delta t)\rangle}{\langle I_A(t)\rangle^2}$ and following its evolution in time.
XPCS allows accessing collective diffusion coefficients\cite{Wagner2006} and investigating the slow dynamics of fluctuating magnetic configurations at the mesoscopic scale with field and temperature\cite{Morley2017}.
Fig.~\ref{XR_figcombined2}f shows $g^2(t)$ measured at the Fe-$L_3$ edge on Fe$_3$O$_4$ MNPs at different temperatures crossing the superparamagnetic blocking transition.


\FloatBarrier

\section{Neutron spin-echo techniques}

\begin{figure*}[!htp]
\centering
\includegraphics[width=1\textwidth]{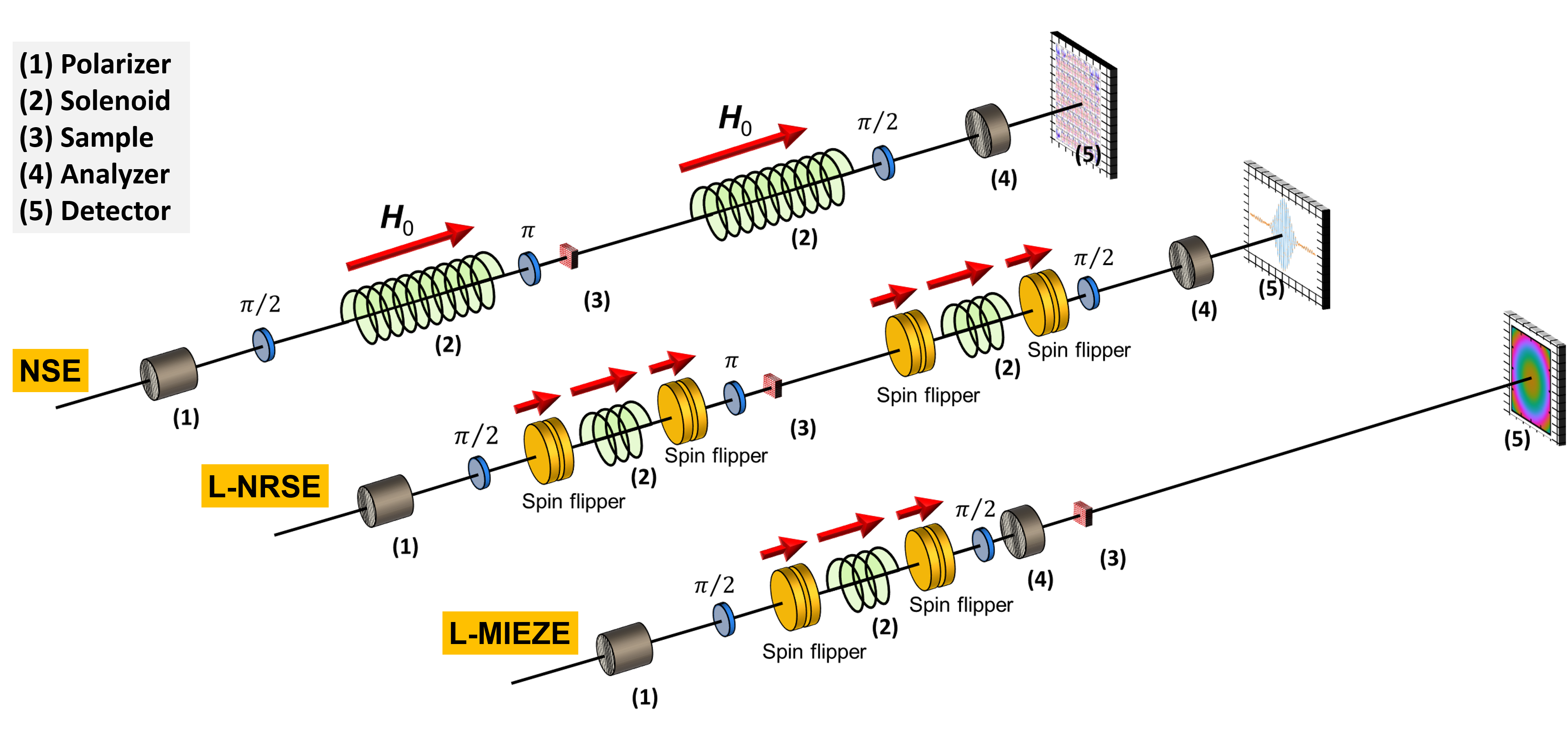}
\caption{Schematic depiction of the spin manipulation components for a classical neutron spin-echo spectrometer (NSE), longitudinal field neutron resonance spin-echo (L-NRSE), and longitudinal modulation of intensity with zero effort (L-MIEZE) from left to right. The red arrow indicates the orientation of the guide field that maintains the neutron polarization along the neutron beam path from left to right. The upstream $\pi/2$ flipper starts the neutron spin precession in the plane perpendicular to the guide field.
For NSE, the $\pi$ flipper reverses the spin states such that for an empty beam the initial polarization is recovered at the second $\pi/2$ flipper, which projects the neutron spin back along the field direction to measure the resulting beam polarization with a polarization analyzer and a neutron detector.
For NRSE and MIEZE, the radio-frequency (RF) spin flippers produce magnetic fields rotating with neutron Larmor frequency that (de)accelerates neutrons yielding a spin-echo on the detector.
}
\label{SE_fig1}
\end{figure*}

\begin{figure*}[ht!]
\centering
\includegraphics[width=1.0\textwidth]{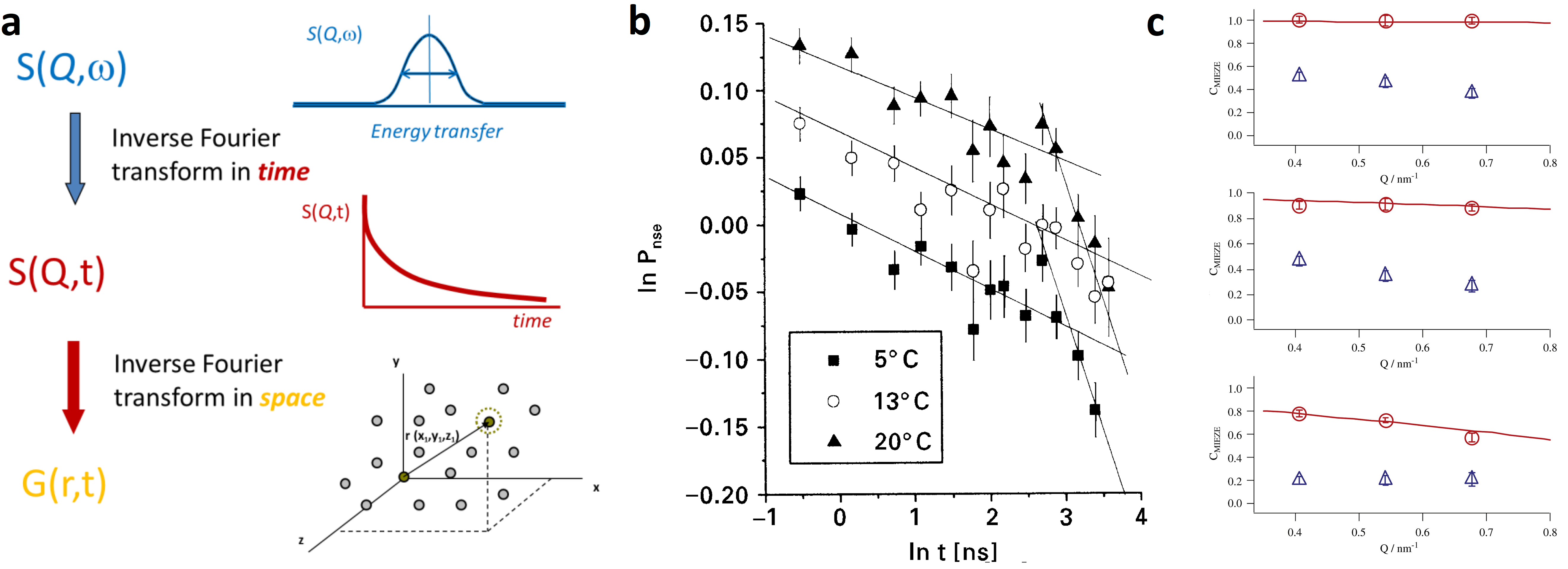}
\caption{(a) With time-of-flight and backscattering spectrometers the dynamic structure factor $S(q,\omega)$ is detected, which can be converted by an energy Fourier transform into the intermediate scattering function $S(q,t)$, and which is directly measured by NSE. The time-dependent self-correlation function $G(r,t)$ follows from a Fourier transform in space. With MIEZE the contrast $C_{MIEZE}\propto\int S(q,\omega)\mathrm{cos}(\omega\tau_{MIEZE})\mathrm{d}\omega=S(q,\tau_{MIEZE})$ is measured for a given MIEZE time $\tau_{MIEZE}$. Adapted from \citet{Arrighi2020} under terms of the CC-BY license.
(b) \citet{Lebedev1999} determined the intermediate scattering function $S(q,t)$ (here called $P_{NSE}$) for a ferrofluid with 10-nm MnZn-ferrite MNPs at different temperatures. The NSE signal in double logarithmic scale nicely shows the existence of two distinct relaxation processes that scale with temperature. Copyright \copyright 1999 Published by Elsevier B.V. 
(c) \citet{Hayashida2009} investigated the relaxation dynamics of 10-nm magnetite MNPs in a ferrofluid by measuring the MIEZE contrast $C_{MIEZE}$ of the nuclear (circles) and magnetic (triangles) scattering at three different $q$-values for three different MIEZE times (from top to bottom: 0.3\,ns, 0.8\,ns, 1.6\,ns). The red line is the instrumental resolution function. The difference between nuclear and magnetic contrast can be attributed to the superparamagnetic relaxation of the MNPs. Copyright \copyright 2008 Elsevier B.V. All rights reserved. 
}
\label{SE_fig2}
\end{figure*}

Neutron spin-echo (NSE) spectroscopy is an ultra-high energy resolution technique, offering energy resolutions better than 1\,neV \cite{JNSE, Farago2015} by encoding the generalized velocities of a neutron with the precession angle (Larmor labeling) in a magnetic field.
Since the development of the first NSE instrument in the 1970s, several more techniques have been established that utilize the neutron spin precession to address the scattering process with unique energy and momentum transfer resolution.
Amongst these are neutron resonant spin-echo (NRSE) and modulation of intensity with zero effort (MIEZE) that utilize spin flippers in resonance with the precessional frequency of a neutron instead of large solenoids used in classical NSE.\cite{Mezei1972} 
Fig.~\ref{SE_fig1} shows schematically the three different NSE spectrometers. 
While classical NSE still holds the records for reaching the highest temporal correlations (and therefore highest energy resolution), NRSE offers the possibility to reach shorter Fourier times with the use of a field subtraction coil.\cite{Jochum2019} 
The MIEZE technique on the other hand is best suited for measurements of magnetic samples as well as measurements requiring an applied magnetic field or other depolarizing sample conditions as all spin manipulation takes place exclusively before the sample region.
NSE techniques measure the intermediate scattering function $S(q,t)$ directly, while other inelastic neutron scattering techniques such as backscattering and time-of-flight spectroscopy (see. Fig.~\ref{SE_fig2}(a)) measure its Fourier transform, the dynamic structure factor $S(q,\omega)$.
Typical timescales accessible with spin-echo techniques are $2\,\mathrm{ps}-50\,\mathrm{ns}$ (NSE) and $0.1\,\mathrm{ps}-20\,\mathrm{ns}$ (MIEZE).

First and foremost, NSE techniques have been used to study diffusion and relaxation processes of MNPs in dispersion as ferrofluids, colloids, or ferrogels, and embedded in a metallic matrix. 
Ferrofluids with different concentrations of magnetite MNPs (size range 3-5\,nm) measured with classical NSE, showed that the energy transfer of the nuclear scattering scales with $q^4$, rather than with $q^2$, which is typical for translational diffusion.\cite{Lebedev1993} 
The authors explained this soft mode through dipole-bond fractals, which break by applying a magnetic field.
The magnetic subsystem is superparamagnetic at low fields and the spin dynamics seem to follow a N\'eel relaxation process. Larger fields induce stronger fluctuations at small q.\cite{Lebedev1993} 
The same group employed a mixture of isotopic $^{54}$Fe and natural magnetite MNPs (size $\approx$ 30\,nm) to eliminate the contribution of the particle pair correlation function due to the varying scattering amplitudes.\cite{Lebedev1999a}
The self-scattering of individual particles reveals a stretched relaxation at low $q$. 
The pair-correlation, however, shows an oscillation-like behavior at large q and a mixed motion of oscillation and diffusion at small $q$. 
The period of oscillation (for both $q$) was found to be 40\,ns with an amplitude of 10\,\AA, which is of the order of magnitude of the length of the surfactant molecule.\cite{Lebedev1999a} 
Similar behavior was found for a magnetite ferrofluid (MNP size $\approx$ 10\,nm) and a magnetite ferrogel containing the same amount of MNPs.\cite{Torok2000}
The dynamics can be described as a motion of chain-like structures composed of random dipole bonds between particles.
For an aqueous ferrofluid constituted of maghemite MNPs (size = 9.8\,nm) with a globally repulsive interparticle potential it was found that for low $q$ the collective diffusion is accelerated due to the repulsive interactions between the particles. 
When a field is applied, a small anisotropy in the diffusion coefficient is predicted by the simulation. 
This is however not observed in the NSE measurements either due to polydispersity, or the relatively small dipolar interaction.\cite{Meriguet2006} 
\citet{Lebedev1999} have studied the diffusion of single-domain particles of MnZn-ferrite (size $\sim$ 10\,nm) in dodecane.
As presented in Fig.~\ref{SE_fig2}b, the temperature-dependent intermediate scattering function nicely shows the existence of two distinct relaxation processes, which can be attributed to a fast and slow diffusion process.
This example shows the strength of NSE techniques for the characterization of a colloidal suspension of MNPs (such as ferrofluids) as relaxation processes can be determined at different $q$-values over a large timescale (several orders of magnitude).
For ferromagnetic Fe(Cu) nanoparticles embedded in a silver matrix of a heterogeneous alloy, NSE provides direct evidence of the interparticle correlations 
of correlated interparticle magnetic relaxation.\cite{Barguin2007}

For magnetic measurements, classical NSE is complicated as the sample region needs to be decoupled from the precession field areas to conserve the Larmor labeling between the two spectrometer arms.  
Since all spin manipulation occurs before the sample position in MIEZE it is ideally suited for measurements of samples under magnetic fields. 
The MIEZE method has recently been developed at various institutes around the world, including the Reactor Institute Delft, the ISIS neutron and muon source,\cite{Geerits2019} Oak Ridge National Laboratory (ORNL),\cite{Brandl2012,Dadisman2020}, with several instruments going into user operation (RESEDA, FRM-II\cite{Franz2019, Franz2019a} VIN-ROSE, J-PARC \cite{ Kawabata2006, Hino2013}).
First MIEZE measurements, studying the dynamic magnetic properties of MNPs, have been performed on a magnetite ferrofluid (MNP size $\approx$ 10\,nm). 
\citet{Hayashida2009} found a superparamagnetic relaxation time of 1.6\,ns for these MNPs.    
This study shows that MIEZE-SANS can be employed to detect directly the moment fluctuations of MNPs.


MIEZE can further play a crucial role in the investigation of magnons in MNP systems.
Such collective magnetic excitations were predicted by Krawczyk and Puszkarski \cite{Krawczyk2008} and have first been experimentally detected by \citet{Krycka2018}, using the triple-axis spectrometer BT7 (NIST \cite{Lynn2012}). 
\citet{Tartakovskaya2008} describe theoretically the dispersion of spin waves in two- and three-dimensional MNP systems by calculations based on a linear combination of atomic orbitals (LCAO) model taking into account dipolar interactions. They predict a phase transition that is caused by the competition between dipolar interaction and uniaxial anisotropy and formulated an expression for the relaxation times and their relation to the dimensionality of a MNP system.
 MIEZE is an excellent tool for studying magnetic excitations as demonstrated on RESEDA (MLZ) on magnons in bulk Fe\cite{Saeubert2019, Kindervater2017} and on fluctuating skyrmion lattices\cite{kindervater2019weak}.

One advantage of NSE techniques is that the energy resolution is decoupled from the width of the used wavelength band.
Therefore, a wavelength selector with a broad wavelength band ($\frac{d\lambda}{\lambda}=10\%$) can be used instead of a monochromator (as is used for example at triple-axis spectrometers) providing a much higher incoming flux onto the sample, without sacrificing energy resolution.
Furthermore, the large dynamic range ($\sim 1\,\mathrm{ps}-10\,\mathrm{ns}$) and the availability of a small angle scattering option, make MIEZE an interesting candidate to study intra- as well as interparticle dynamics, like the relaxation processes at the interfaces between particle core and shell.\cite{Oberdick2018} 
In dispersion, these effects are obscured by the presence of Brownian diffusion. 
Hence, to investigate the moment dynamics one should aim to embed the MNPs in a hard matrix but that is rather transparent to neutrons, such as Aluminium or Zirconium.
On the other hand, studying the diffusion and relaxation of MNPs in viscoelastic matrices can be an interesting research field in itself in particular concerning biomedical applications (e.g., regarding the interaction of MNPs with physiological/cellular environments) or soft actuators and robotics (e.g., regarding the binding of MNPs in ferrogels and elastomers).



\FloatBarrier

\section{Micromagnetic simulations}\label{MicroMag}

\begin{figure}[!hb]
\centering
\includegraphics[width=0.49\textwidth]{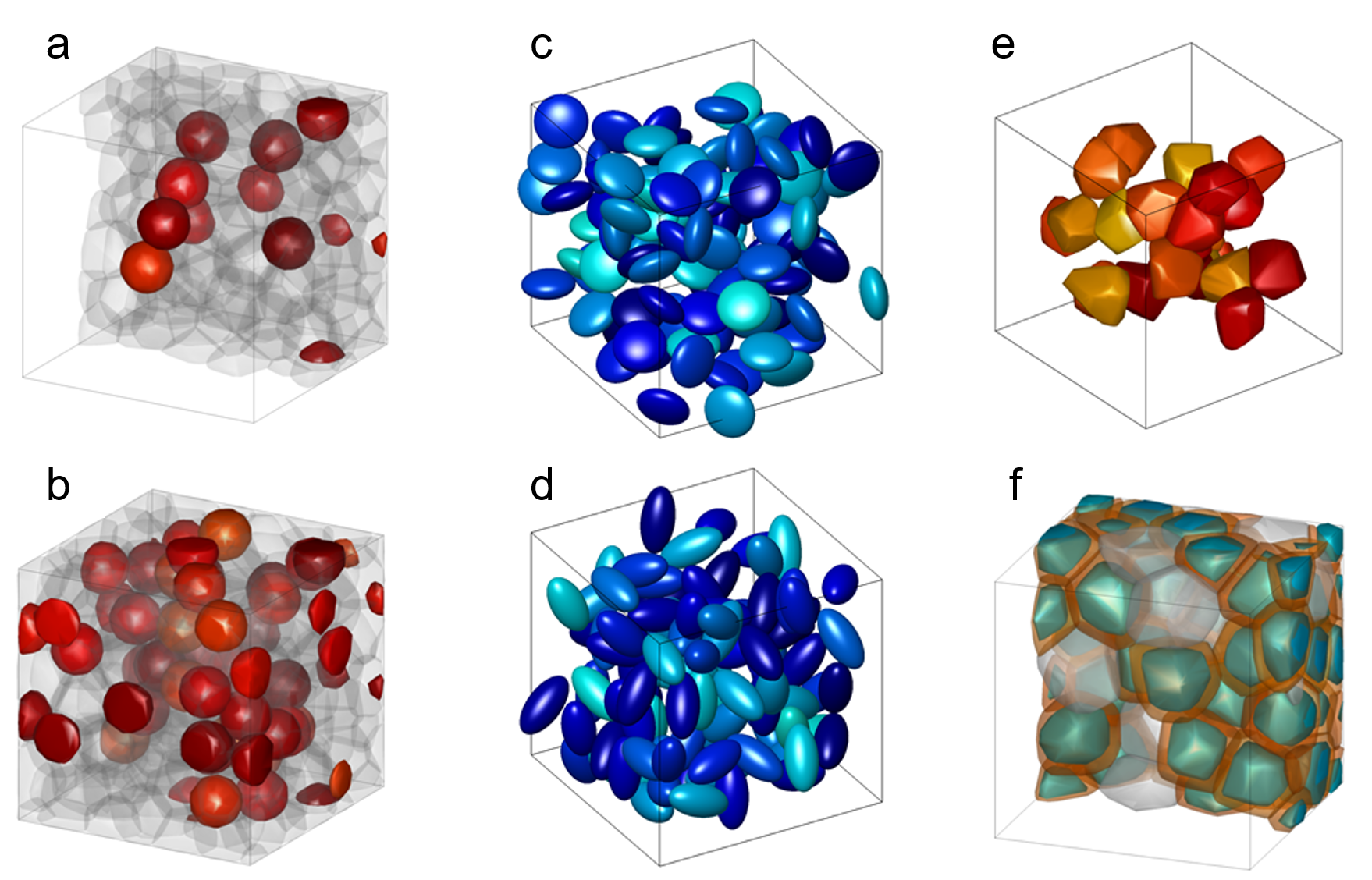}
\caption{Examples for the spatial distribution of nanoparticles with various shapes. (a),~(b)~Spherical particles with different volume fractions. Reprinted figures with permission from \citet{Vivas2020}. Copyright 2020 by the
American Physical Society. (c),~(d)~Ellipsoids of revolution with different aspect ratios. (e)~Clustered particles with polyhedron shapes. (f)~A dense system of particles with a core-shell microstructure.}
\label{figmumag:geom}
\end{figure}

\begin{figure*}[h]
\centering                                                \resizebox{0.95\textwidth}{!}{\includegraphics{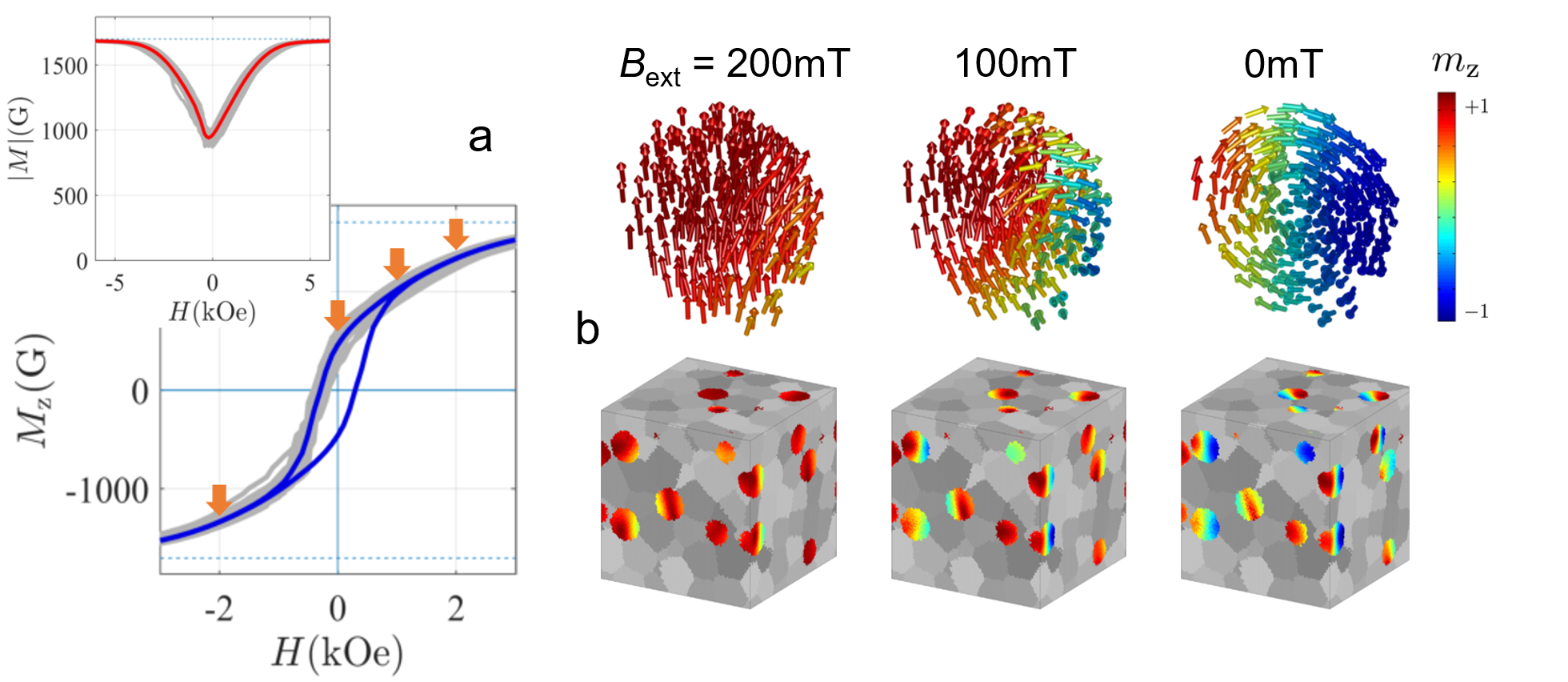}}
\caption{Evolution of the magnetization of spherical Fe nanoparticles with a diameter of $D = 40 \, \rm{nm}$ and a volume fraction of $x_p = 15 \%$. (a)~Hysteresis loop (blue line) averaged over many geometrical realizations (gray lines); inset:~$|M|(B)$ magnetization dependence (red line) showing the deviation of the particle's magnetization from the single-domain state (dashed blue line in the inset corresponds to the saturation magnetization of iron); (b)~Magnetization states at different fields for an individual particle (upper row) and over the whole simulation box (lower row).}
\label{figmumag:magn}
\end{figure*}

\begin{figure}[h]
\centering                                                \resizebox{0.5\textwidth}{!}{\includegraphics{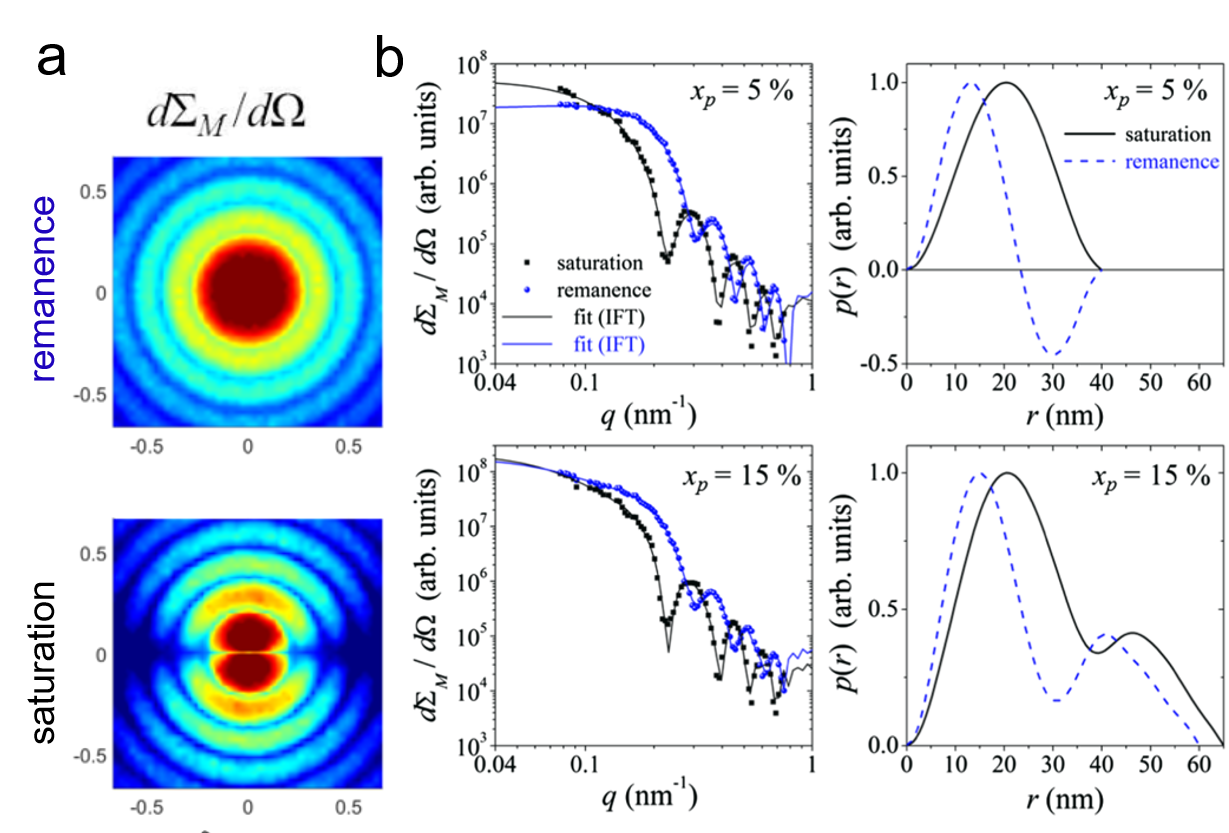}}
\caption{Results of the micromagnetic SANS modelling for an ensemble of spherical Fe particles with a diameter of $D = 40 \, \rm{nm}$. (a)~The total magnetic SANS cross-section in the perpendicular scattering geometry ($x_p = 15 \%$). (b)~Azimutally-averaged total magnetic SANS cross-sections and corresponding distance distribution functions $p(r)$ for $x_p = 5 \%$ and $x_p = 15 \%$.  Reprinted figure with permission from \citet{Vivas2020}. Copyright 2020 by the
American Physical Society.}

\label{figMMcombined}
\end{figure}

A model-based analysis of magnetic scattering data (either in magnetic SANS, neutron reflectometry, GISANS, or XRMS) is feasible only for a few selected sample microstructures (e.g. with a low polydispersity and a well-defined particle morphology). To circumvent this complexity, the application of micromagnetic simulations was proposed recently to \textit{translate} the reciprocal-space data into real space. This indirect approach for data analysis was so far used for magnetic SANS and bears great potential for the other small-angle scattering techniques. Additionally, dynamic micromagnetic simulations can be performed, which can be used in principle to evaluate time-resolved measurements and NSE data.
In the following, we give a brief overview of the state-of-the-art characterization of MNPs using micromagnetic simulations. 

The `classical' (mesoscopic) approach to micromagnetics\cite{Brown-book} predicts the static magnetization configuration or the magnetization dynamics with a spatial resolution of a few nanometers. Commonly, five contributions to the total magnetic Gibbs free energy are taken into account:~the Zeeman energy in the external magnetic field, magnetocrystalline anisotropy, symmetric exchange, the antisymmetric Dzyaloshinskii-Moriya interaction (DMI), and the magnetodipolar energy. On the nanometer length scale, one can ignore the discrete atomic structure and a continuum approximation suffices to resolve the details of typical magnetization structures both in thin film and bulk systems. Finally, the minimization of the total magnetic energy is performed, e.g. by an optimized version of a gradient method employing the dissipation part of the Landau-Lifshitz equation of motion for magnetic moments.\cite{landau35} As for all physical theories, the violation of applicability limits (e.g. attempts to achieve a better resolution than allowed by mesoscopic theories) leads to physically irrelevant results. This remark is in particular important for small-sized magnetic particles, where atomistic modeling is often the more appropriate choice, e.g. to observe the effect of antiferromagnetic bonds across antiphase boundaries within MNPs,\cite{Nedelkoski2017} or DMI-induced canted spins across core/shell Fe$_3$O$_4$/Mn$_x$Fe$_{3-x}$O$_4$ MNPs\cite{Oberdick2018}.


Numerical micromagnetics utilizes two different approaches for spatial discretization, finite-difference (FDM) and finite-element methods (FEM). FDM uses commonly a simple rectangular grid that opens up the possibility to apply fast Fourier transformation (FFT) to compute the long-range magnetodipolar interaction energy. A drawback of FDM consists in the difficulty to adequately represent non-flat surfaces (e.g. the surface of a spherical nanoparticle) so that an increased computational effort is required to describe curved surfaces with a very fine discretization mesh.
A novel micromagnetic approach based on a polyhedral finite-element mesh has been developed by~\citet{Erokhin2012prb1}. This enables the study of the magnetization distribution of a wide range of mono- and polycrystalline materials as well as MNPs embedded in a nonmagnetic matrix. Moreover, the approach allows for a highly flexible mesh generation combined with an efficient FFT-based magnetodipolar energy calculation. The methodology was successfully applied to explain the details of the magnetization reversal in Fe-based soft magnetic alloys,\cite{Michels2014jmmm} Nd-Fe-B nanocomposites possessing a core-shell grain structure,\cite{Erokhin2018} and to optimize the structural properties of permanent-magnet materials based on ferrite alloys.\cite{Erokhin2017} Fig.~\ref{figmumag:geom} presents examples of MNP systems that can be efficiently discretized by a polyhedron mesh keeping the total number of elements at a computationally reasonable level. These examples include spherical particle systems with different volume fractions, a gradual transition from oblate to prolate particle shapes, clustered particles of polyhedron shapes, and core-shell microstructures.

In the following, we discuss the magnetic SANS response of an ensemble of spherical, defect-free Fe MNPs. For such a system, the magnetization reversal process occurs exclusively via the coherent magnetization rotation for particle sizes below the critical single-domain size ($D_{\rm cr} = 15 \, \rm{nm}$ for Fe.\cite{Vivas2020}) Only for a dilute ensemble of particles with diameters smaller than $D_{\rm cr}$ and having a cubic magnetic anisotropy, one can expect that the magnetization behavior resembles the analytical result for randomly-oriented Stoner-Wohlfarth particles;\cite{Stoner1948} for all other cases, micromagnetic modeling is required. As an example, Fig.~\ref{figmumag:magn} depicts the magnetization reversal of Fe MNPs with a diameter of $D = 40 \, \rm{nm}$ and a magnetic volume fraction of $x_p = 15 \%$. The hysteresis curve (Fig.~\ref{figmumag:magn}, blue line) shows the typical behavior of a soft magnetic material, which starts to demagnetize in positive external fields due to the influence of the magnetodipolar interaction. This influence has a twofold impact:~a strong self-generated demagnetization of the particle and a weaker interparticle coupling. To quantify the average magnetization state of the particles as a function of an applied field, the absolute magnetization dependence $|M| = N_p^{-1} \sum_{i=1}^{N_p} |M_i|$ (red line) can be used, where
\begin{equation}
\label{eq:mdeviation}
|M_i| = \sqrt{(M_{i,x})^2+(M_{i,y})^2+(M_{i,z})^2}
\end{equation}
is the total magnetization of the $i$th-particle, and $N_p$ denotes the number of particles in the simulation volume. Therefore, the quantity $|M|-M_s$ describes the average deviation of the particle's magnetization distribution from the single-domain state. In particular, the minimum of the $|M|(B)$-dependence corresponds to a highly nonuniform magnetization state. As shown in Fig.~\ref{figmumag:magn}b, this state corresponds to a vortex-type spin configuration, which is predominant at remanence.

With the real-space magnetization configurations obtained from micromagnetic simulations, one can directly compare 2D SANS patterns from modeling and experiment. This is possible because both techniques address the same resolution range ($\sim 1-100 \, \rm{nm}$). In magnetic neutron scattering, one measures a weighted sum of magnetization components, and the numerical simulations allow one to decrypt the response of individual magnetic contributions and to study the effect of variations in microstructure and magnetic parameters.\cite{Erokhin2012prb2} This fact renders numerical micromagnetics the go-to tool to attain a comprehensive SANS data interpretation of 3D magnetic structures.

Fig.~\ref{figMMcombined}a represents the results of such a calculation for an ensemble of spherical $40$-nm-sized Fe particles. Shown is the total magnetic SANS cross-section at remanence and at saturation ($x_p = 15 \%$). While at large fields the main contribution to the total scattering originates from the longitudinal magnetization component $\widetilde{M}_z$, at remanence all Fourier components play an important role in the formation of the resulting SANS pattern. Nonzero transversal magnetization components $\widetilde{M}_{x,y}$ at small momentum-transfer vectors indicate interparticle correlations, while for $q$-ranges associated to the particle interior it indicates an inhomogeneous spin structure and eventually the formation of a multidomain state.
As was shown in \citet{Vivas2020}, the radially-averaged total SANS cross-section can be used for the further analysis of the magnetization configuration. For example, its Fourier transformation to the so-called pair-distance distribution function demonstrates a quantitative different behavior at saturation and in the remanent state. At saturation, the $p(r)$ for almost homogeneously magnetized spheres in the dilute limit ($x_p = 5 \%$) coincides with the analytical solution, whereas the vortex-like configurations at remanence produce an oscillating pair-distance distribution function (Fig.~\ref{figMMcombined}b, upper row). In this way, it is seen that the $p(r)$-representation has an advantage over the azimuthally-averaged total magnetic SANS cross-section due to the much larger sensitivity of $p(r)$ to the details of the magnetization configuration.
Applying the same approach to the relatively dense ensemble ($x_p = 15 \%$) of spherical Fe particles of the same size, we observe another distinctive feature (Fig.~\ref{figMMcombined}b, lower row). Both (saturated and remanent) curves indicate a second maximum due to the increased influence of the interparticle magnetodipolar interaction. The position of the first local maximum in the saturated regime coincides with the $x_p = 5 \%$ case, which is an indication for the fully magnetized state of spheres of any volume fraction. The peak shift from saturation to remanence, as for the dilute case, is again attributed to an inhomogeneous magnetization configuration of the particles.

The above example suggests that magnetic SANS is highly sensitive to the internal magnetization profile of MNPs. Thus, micromagnetic simulations are a valuable tool to analyze experimental data, as shown in \citet{Bersweiler2019}, where micromagnetic simulations were employed to interpret measured magnetic SANS profiles. The combination of simulations and experiments confirmed that MnZn-ferrite MNPs with diameters of around 10\,nm are in a homogeneously magnetized single-domain state with increasing diameter the magnetization configuration deviates from the collinear alignment. For the analysis of nuclear small-angle scattering data of biomacromolecules, the usage of ab initio\cite{svergun2002advances} and molecular dynamics simulations\cite{boldon2015review} are already well established for several years. Considering that the capabilities of micromagnetic modeling have dramatically increased in recent years thanks to the deployment of graphical processing units (GPU), which speed up calculations considerably,\cite{leliaert2018fast} we believe that also for the analysis of magnetic small-angle scattering data micromagnetic simulations will become the standard approach.

\FloatBarrier


\section{Summary and perspectives}

Small-angle scattering of X-rays and neutrons is sensitive to the chemical composition and magnetization profile of nanostructured samples on a mesoscopic length scale from about 1 to a few hundred nanometers.
The strength of small-angle scattering is that it can be employed to directly interrelate the macroscopic behavior and magnetic properties with their nuclear-magnetic nano-/microstructure.
As shown and discussed for various examples, such a multiscale characterization can help to optimize the MNP systems for specific applications.
Additionally, advanced small-angle scattering techniques exist that allow time-resolved experiments.
Such dynamic measurements are especially useful to monitor time-dependent synthesis processes, or they allow for the characterization of MNPs intended for applications where the relaxation dynamics of the system play a decisive role.

Despite small-angle scattering being such a powerful tool for MNP characterization, it is significantly underused by the MNP community, which may be attributed to its reputation for being difficult and inaccessible as the scattering signal is presented in reciprocal space units.
In contrast to other, localized characterization techniques, however, scattering assesses the genuine compositional and structural arrangement averaged over many large sample volumes.
The data analysis is highly sensitive to, e.g. size distributions that smear out characteristic features of the signal expected from individual, perfect shapes. This is an advantage as it allows obtaining immediate information regarding distribution functions and dispersity of MNP ensembles from a single measurement.
As a result, SAXS – which laboratory instruments perform routinely nowadays – is predestined for determining particle size distributions of MNPs and has become the standard approach for a pre-characterization of commercial MNP suspensions.
To extract the size distributions, the particles need to have a well-defined shape so that the intensity can be fitted with the appropriate model function.
In the case of less well-defined particles, model-independent approaches can be used to reveal the autocorrelation functions of the scattering length density profile from the SAXS data by inverse Fourier transforms, which allows gaining valuable information about the average particle morphology and density profile.

Similar to SAXS and nuclear SANS, magnetic SANS data are either analyzed in reciprocal space via model fits (at least for well-defined and nearly monodisperse samples) or in real space by applying inverse Fourier transforms.
Magnetic neutron scattering is very sensitive to differences in the magnetization distribution and allows observing weak disorder superposed on the average magnetization.
However, magnetic scattering only sees the magnetization density perpendicular to the scattering vector.
Thus, the extracted correlation functions do not result from a scalar composition profile (as for nuclear scattering).
The magnetic correlation function measures the variation in the magnetization vector field over distance while considering spatial orientations\cite{Roth2018}, which makes their interpretation in terms of model functions challenging.

To directly relate data to physically motivated magnetization configurations, micromagnetic simulations have emerged as a tool to interpret magnetic SANS experiments. 
This is a new research avenue that will gain steam in upcoming years thanks to the ever-increasing speed\cite{leliaert2018fast} and accessibility of open-source simulation tools\cite{leliaert2019tomorrow}.
It is easy to imagine that also neutron reflectometry, GISANS, and also XRMS will benefit from a combined data analysis with micromagnetic simulations in upcoming years.
In general, it can be said that improved tools for data analysis (such as software and simulations) are key to making small-angle scattering techniques more accessible for non-expert users.
In particular, micromagnetic simulations have the huge advantage of basically translating the often non-intuitive reciprocal data into comprehensible real-space images.
However, micromagnetics assumes a continuous magnetization, and thus to simulate the internal magnetization profile of fine particles, e.g., to elucidate the influence of structural defects, spin coupling across an interface, and magnetic frustration, atomistic simulations may be better suited.\cite{Oberdick2018}
Regarding the calculation of interparticle correlations, on the other hand, to resolve short- and long-range order in MNP assemblies Monte-Carlo simulations can be employed as demonstrated recently for artificial spin ice.\cite{pip2021direct}
For data fitting, improved tools are developed such as Bayesian approaches for improving model fits of neutron reflectometry\cite{mccluskey2020general} and SANS\cite{bersweiler2020benefits} data to increase the predictive power of the fits and maximize the information density.
Similar to model-fits also inverse Fourier transforms are usually ill-posed problems, and thus new approaches have been introduced in recent years that improve data analysis either by applying Bayesian analysis to find the most probable solution\cite{vestergaard2006application} or by using novel approaches such as the singular value decomposition\cite{bender2019using} or fast iterative method to reveal the real-space 2D correlations\cite{bender2021unraveling}.
In this context, it is safe to assume that also machine learning will massively contribute in upcoming years. 
In fact, for neutron reflectometry first, approaches are published which aim to achieve an automatized data analysis to extract parameters from scattering data without expert knowledge.\cite{mironov2021towards,loaiza2021towards}
This concept can be easily transferred to other scattering techniques such as SANS to classify the most appropriate model.\cite{Archibald2020}

Regarding future experiments, we think that combining small-angle scattering experiments with micromagnetic simulations could be particularly useful to investigate exotic nanoparticles such as hollow particles\cite{khurshid2021hollow} including  nanorings,\cite{das2020magnetic} nanotubes,\cite{landeros2007reversal} and other shape-anisotropic hollow particles,\cite{niraula2021engineering} or nanodots\cite{goiriena2017magnetization} and nano-octopods\cite{garnero2021single}.
Recently, it was shown that the coercivity of MNPs is enhanced by the exchange coupling at the interface of ferrimagnetic and antiferromagnetic self-assembled monolayers.\cite{sawano2020enhancement}
For this and similar systems, reflectometry, GISANS, and XRMS can reveal the correlation between structural and magnetic ordering between the layers over a large size range with nanometer resolution.
Micromagnetic simulations allow linking the microscopic structure to the corresponding macroscopic magnetic properties. Thus, these studies are a powerful approach to optimize the particle properties for specific applications.

In addition to static simulations, dynamic micromagnetic simulations can be performed, which will be invaluable for the analysis of time-resolved data.
So far, most stroboscopic small-angle scattering measurements of MNPs were performed to monitor structural changes and the diffusional motion of nanoparticles.
Such experiments can be easily extended to study clustering dynamics and in combination with dynamic micromagnetic simulations could be a great approach to optimize the structural and magnetic properties of supraparticles\cite{guo2013magnetic} or particle clusters\cite{peng2018superparamagnetic}.
The big strength of stroboscopic small-angle scattering techniques (e.g. SANS or coherent-XRMS) is that they can also reveal the internal magnetization dynamics of the MNPs.
Thanks to the high spatial resolution, stroboscopic small-angle scattering techniques have the potential to access unique information regarding the relaxation dynamics of MNP samples when combined with micromagnetic simulations.
Such experiments enable to study structure formation of MNPs in alternating magnetic fields, similar to what is already done in static fields, or resolve the time-modulated internal magnetization profile of MNPs to compare it with their static profile.
By combining simultaneous magnetic hyperthermia experiments with small-angle scattering, a\cite{carlton2020situ} fundamentally new understanding of the microscopic dynamics governing magnetic heating could be obtained and systematic studies (varying the particles sizes, morphology, etc.) will help to optimize the particle properties for hyperthermia applications.
These simultaneous, in-situ measurements would further give insight into the local modifications while heating MNPs\cite{cazares2019recent} and catalytic applications such as CO$_2$ hydrogenation\cite{bordet2016magnetically} and electrolysis\cite{niether2018improved}.

In this context, we also believe that NSE and here especially MIEZE-SANS will gain significant popularity for MNP characterization once the methodology is better established.
With MIEZE, spin dynamics in the picosecond to nanosecond regime are accessible.
This time range is especially interesting to study transversal, intra-well moment relaxation in MNPs.\cite{neugebauer2020investigation}
Recent experiments revealed that the magnetic heating is significantly increased for some samples compared to classical magnetic hyperthermia with GHz-frequencies fields and low field amplitudes of around 0.2 mT.\cite{lee2021ultra}
The extremely high heating rates of $~ 150$\,K/s were achieved when the excitation frequency matched the characteristic transversal relaxation frequency of the particle moments.
Additionally, within densely packed MNP assemblies, spin-waves may form due to a dipolar coupling between the precessing spins.\cite{Krycka2018} The expected time and length scales are accessible with MIEZE-SANS and should be investigated similar to studies performed on bulk ferromagnets.\cite{Saeubert2019}

Finally, we want to mention another small-angle scattering technique that may be useful for future MNP studies. Similar to NSE, which encodes changes in neutron velocity due to energy transfer, Spin-echo SANS (SESANS) utilizes the Larmor precession of a neutron to resolve minute scattering angles.
SESANS allows observing microstructures in a single measurement over three orders of magnitude from nm to several $\mu\,m$, well beyond the conventional SANS regime.
It has been successfully used to study the structural properties of non-magnetic NP systems, such as porous silica particles.\cite{parnell2016porosity} SESANS can also be applied to magnetic samples with sufficient magnetic scattering contrast.\cite{Grigoriev2006}
SEMSANS, a comparable approach to MIEZE with neutron spin manipulation before the sample, enables the measurement of spatial magnetic correlations with flexible and varying magnetic field configurations.\cite{Li2021}
The technique could be beneficial to characterize large MNPs and dense assemblies thereof, including magnetic superstructures, and to study their field-dependent response.



\section{Author contributions}
DH and PB conceived the idea, devised the project, and drafted the manuscript.
PB wrote the introduction, and summary, and perspectives.
MB contributed the SasView computations and designed the artwork.
SE and DB wrote the section on micromagnetic simulation.
KC wrote the overview on X-ray magnetic scattering and spectroscopy.
DAV contributed with DH to the sections on small-angle X-ray and neutron scattering and time-resolved in-situ measurements.
AQ and SD authored the part on Reflectometry and grazing-incidence scattering.
JKJ took the lead on neutron spin-echo techniques.
AM provided critical feedback on all techniques and the micromagnetic simulations.
All authors critically read and discussed the manuscript.

 \section*{Acknowledgments}
This work benefited from the use of the SasView application, originally developed under NSF award DMR-0520547. SasView also contains code developed with funding from the European Union’s Horizon 2020 research and innovation program under the SINE2020 project, grant agreement No 654000. Andreas Michels and Mathias Bersweiler acknowledge financial support from the National Research Fund of Luxembourg. Asma Qdemat and Sabrina Disch acknowledge financial support from the German Research Foundation (DFG Grant No. DI 1788/2-1).



\balance



\FloatBarrier
\bibliography{bib}
\bibliographystyle{rsc}

\end{document}